\documentclass[12pt,english,longbibliography,nofootinbib,superscriptaddress,12pt,sort&compress,showkeys]{revtex4-1}
\usepackage[T1]{fontenc}
\usepackage[latin9]{inputenc}
\usepackage{xcolor}
\usepackage{array}
\usepackage{longtable}
\usepackage{amsmath}
\usepackage{amssymb}
\usepackage{graphicx}
\usepackage{esint}
\PassOptionsToPackage{normalem}{ulem}
\usepackage{ulem}

\makeatletter

\providecommand{\tabularnewline}{\\}
\providecolor{lyxadded}{rgb}{0,0,1}
\providecolor{lyxdeleted}{rgb}{1,0,0}
\DeclareRobustCommand{\lyxadded}[3]{{\color{lyxadded}{}#3}}

\DeclareRobustCommand{\lyxsout}[1]{\ifx\\#1\else\sout{#1}\fi}

\usepackage{indentfirst}%\DeclareGraphicsExtensions{.pdf,.png,.jpg}
\usepackage{amsfonts}
\usepackage[T1]{fontenc}
\usepackage{ae,aecompl}
\usepackage{sidecap}
\usepackage[section]{placeins}
\usepackage{epsf}
\makeatother

\usepackage{babel}
\date{\today}

\usepackage{chngcntr}

\makeatother

\usepackage{babel}
\begin{document}
\title{Effect of Vacancy Creation and Annihilation on Grain Boundary Motion}
\author{\noindent G. B. McFadden}
\address{\noindent Information Technology Laboratory, National Institute of
Standards and Technology, Gaithersburg, Maryland 20899, USA}
\author{\noindent W. J. Boettinger}
\address{\noindent Materials Science and Engineering Division, National Institute
of Standards and Technology, Gaithersburg, Maryland 20899, USA}
\author{\noindent Y. Mishin}
\address{Department of Physics and Astronomy, MSN 3F3, George Mason University,
Fairfax, Virginia 22030, USA}
\begin{abstract}
Interaction of vacancies with grain boundaries (GBs) is involved in
many processes occurring in materials, including radiation damage
healing, diffusional creep, and solid-state sintering. We analyze a model describing a set of
processes occurring at a GB in the presence of a non-equilibrium,
non-homogeneous vacancy concentration. Such processes include vacancy
diffusion toward, away from, and across the GB, vacancy generation and absorption at the
GB, and GB migration. Numerical calculations within this model reveal
that the coupling among the different processes gives rise to interesting
phenomena, such as vacancy-driven GB motion and accelerated vacancy
generation/absorption due to GB motion. The key combinations of the
model parameters that control the kinetic regimes of the vacancy-GB
interactions are identified via a linear stability analysis. Possible
applications and extensions of the model are discussed.
\end{abstract}
\maketitle

\section{Introduction}

Many processes in materials involve vacancy generation or absorption
by grain boundaries (GBs). For example, in materials subjected to
energetic radiation, large amounts of vacancies and interstitials
are produced from displacement cascades, which then diffuse to dislocations
and GBs and are absorbed by them with a varying
degree of efficiency. Vacancy absorption can cause dislocation climb
and displacements of GBs. In turn, a moving GB can sweep out a larger
amount of vacancies than a stationary one, which can enhance its efficiency
as a vacancy sink. Another example is furnished by the creep deformation
of materials. At high temperatures and under sustained mechanical
loads, many materials undergo slow, time-dependent plastic flow controlled
by diffusion of vacancies between sources and sinks located at GBs
\citep{Nabarro1948,Herring1950,Coble,Farghalli01,Svoboda2006,Svoboda2011a,Mishin-2013}.
For example, in the case of tensile creep, the deformation occurs
by elongation of the grains in the tensile direction by vacancy generation
at GBs normal to that direction and vacancy absorption at GBs parallel
to that direction.
Interaction of vacancies with moving grain boundaries is also an important
process during solid-state sintering \citep{Herring:1951aa,Herring53}.
Understanding
such interactions is important for the development of materials with
improved radiation tolerance, creep-resistance and other desirable
service characteristics.

In a previous paper \citep{Mishin:2015ad}, an irreversible thermodynamics
theory of creep deformation of polycrystalline materials was developed
based on a sharp-interface treatment of GBs. The GBs were represented
by geometric surfaces capable of vacancy generation and absorption
and moving under local thermodynamic forces. Kinetic equations of
creep deformation were derived taking into account 
capillary effects, deviations of the vacancy concentration from equilibrium,
and the effect of mechanical stresses on GB thermodynamics and kinetics.
For future applications, the general equations of the theory were
specialized for the particular case of a linear-elastic solid with
a small vacancy concentration. 

In the present paper we apply the theory \citep{Mishin:2015ad} to
study the particular case of a single planar GB separating two semi-infinite
grains in a single-component system, with the goal of gaining a more
detailed understanding of the GB-vacancy interactions. As the initial
condition, a non-equilibrium vacancy distribution is created in both
grains. This initial state can be thought of as produced by irradiation
of the material by energetic particles.
Radiation cascades produce both vacancies and interstitials. Interstitials
are much more mobile than vacancies and soon annihilate at various
defects or form clusters. In this work we neglect the role of the
interstitial clusters and focus the attention on later stages of the
system evolution which are controlled by vacancy diffusion towards
grain boundaries. Alternatively, the initial state could be obtained
by rapidly quenching the material from a higher temperature or rapidly
heating it up to a higher temperature.
Non-equilibrium vacancies can also be created by vacancy fluxes driven
by differences between the local equilibrium vacancy concentrations
near surfaces with different curvature. This process is relevant to
solid-state sintering and has recently been studied by the phase-field
approach \citep{Kazaryan:1999aa,Wang:2006aa,Abdeljawad:2019aa} and
other computational methods \citep{Li-2011}.
Our model assumes
that the vacancy sinks and sources only exist at the GB. Agglomeration
of vacancies into clusters inside the grains is neglected. The operation
of sinks and sources at the GB drives the system toward thermodynamic
equilibrium by sucking in the excess vacancies or by injecting additional
vacancies into the grains. In addition to the vacancy generation and
annihilation at the boundary, the equilibration (relaxation) process
involves vacancy diffusion toward or away from the boundary, diffusion
of atoms across the boundary, and GB migration driven by the discontinuity
in the vacancy concentration that arises across the boundary. We perform
a detailed parametric analysis of the vacancy relaxation processes,
identifying the possible kinetic regimes and the dimensionless parameters
governing the individual relaxation modes. 

In Section \ref{sec:Theory} we recap the main assumptions of the
theory \citep{Mishin:2015ad} and specialize its equations for the
particular case of a planar GB between two linearly-elastic
grains. In Section \ref{sec:Numerical-study} we perform a numerical
study of the vacancy evolution under various initial conditions and
some of the most representative combinations of the model parameters.
The choices of the model parameters are largely guided by the linear
stability analysis of the model reported in Appendix
B. This analysis identifies the normal modes of the vacancy relaxation
process and the dimensionless combinations of the parameters corresponding
to the normal modes. The numerical calculations demonstrate the effect
of GB motion on the vacancy absorption, the vacancy-driven GB migration,
and other interesting effects arising from the interaction of moving
GB with vacancies. In Section \ref{sec:Conclusions} we summarize
the work and formulate conclusions.
The supplementary file accompanying this paper \citep{Supplementary-Material}
contains a complete set of figures obtained by the numerical calculations.

\section{Theory\label{sec:Theory}}

\subsection{General equations for a planar grain boundary}

\begin{figure}
\begin{centering}
\includegraphics[scale=0.65]{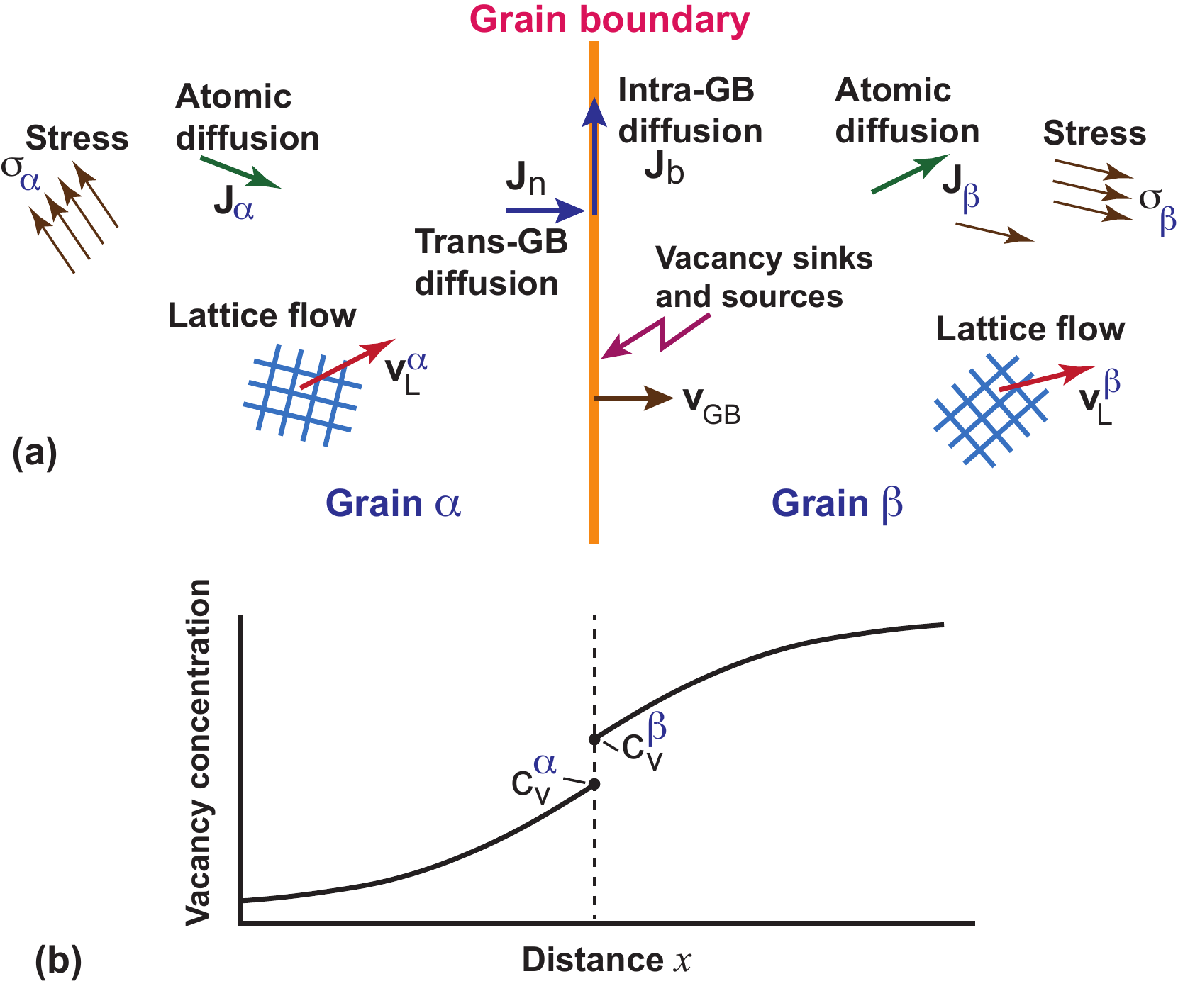}\caption{The physical processes
occurring at the grain boundary and in the grains within the proposed
model.\label{fig:The-physical-processes}}
\par\end{centering}
\end{figure}

Consider two semi-infinite grains in a single-component solid. The
grains are labeled $\alpha$ and $\beta$ and are separated by a planar
GB (Fig.~\ref{fig:The-physical-processes}).
The lattice site generation and annihilation process occurring at
the GB causes rigid-body motion of the lattice of each grain toward
or away from the GB. The grains can also translate past each other
parallel to the GB plane by GB sliding or due to the shear-coupling
effect \citep{Cahn06a,Cahn06b}. The respective velocities of the
grains with respect to a chosen laboratory reference frame are denoted
$\mathbf{v}_{L}^{\alpha}$ and $\mathbf{v}_{L}^{\beta}$, respectively.\footnote{See the Table in
Appendix A for a list of the notation used in this paper.} In addition to the rigid-body motion, atoms in the grains can diffuse
through the lattice by the vacancy mechanism. The atomic diffusion
fluxes relative to the reference frame attached to the lattices of
the grains are $\mathbf{J}_{L}^{\alpha}$ and $\mathbf{J}_{L}^{\beta}$,
respectively. Atoms can also diffuse along and across the moving GB.
The respective diffusion fluxes measured relative to the GB structure
are $\mathbf{J}_{b}$ and $\mathbf{J}_{n}=J_{n}\mathbf{n}^{\alpha}$,
where $\mathbf{n}^{\alpha}$ is the unit normal vector pointing from
grain $\alpha$ into grain $\beta$. The two-dimensional diffusion
flux $\mathbf{J}_{b}$ measures the number of atoms in the GB plane
crossing a unit GB length per unit time.

The entire bicrystal is subject to an applied mechanical stress. The
elastic stresses arising inside the grains are represented by two
Cauchy stress tensors $\mathbf{\boldsymbol{\sigma}_{\alpha}}$ and
$\mathbf{\boldsymbol{\sigma}_{\beta}}$, respectively.

Thermodynamic properties of the interior regions of the grains are
described by the grand-canonical potentials (per unit volume) 
\begin{equation}
\omega_{\alpha}\equiv\dfrac{f_{s}^{\alpha}-M_{\alpha}c_{\alpha}}{\Omega_{\alpha}}\label{eq:1}
\end{equation}
and
\begin{equation}
\omega_{\beta}\equiv\dfrac{f_{s}^{\beta}-M_{\beta}c_{\beta}}{\Omega_{\beta}},\label{eq:2}
\end{equation}
where $f_{s}^{\alpha}$ and $f_{s}^{\beta}$ are the free energies
per lattice site, $\Omega_{\alpha}$ and $\Omega_{\beta}$ are the
respective atomic volumes, $M_{\alpha}=\partial f_{s}^{\alpha}/\partial c_{\alpha}$
and $M_{\beta}=\partial f_{s}^{\beta}/\partial c_{\beta}$ are the
diffusion potentials of atoms relative to the vacancies, and $c_{\alpha}$
and $c_{\beta}$ are the atomic concentrations (fractions of occupied
lattice sites) in the grains. The grand-canonical and diffusion potentials
are functions of temperature, concentration and lattice strain relative
to a chosen reference state. In this work, temperature is assumed
to be uniform and permanently fixed. 

Diffusion inside the grains follows the phenomenological equations
\citep{Mishin:2015ad} 
\begin{equation}
\mathbf{J}_{L}^{\alpha}=-L_{\alpha}\nabla M_{\alpha},\label{eq:3}
\end{equation}
and
\begin{equation}
\mathbf{J}_{L}^{\beta}=-L_{\beta}\nabla M_{\beta},\label{eq:4}
\end{equation}
respectively, where $L_{\alpha}>0$ and $L_{\beta}>0$ are diffusion
kinetic coefficients. Combined with the mass conservation law, these
equations describe the evolution of the atomic concentrations in space
and time. Accordingly, the grand-canonical potentials, diffusion potentials
and stresses are functions of the position vector $\mathbf{x}$ and
time $t$.

For the description of GB processes, we will use the values of the
aforementioned functions extrapolated toward the GB from either grain.
The following notation is introduced for the extrapolated properties
\citep{Mishin:2015ad}. For any vector field $\mathbf{A}(\mathbf{x},t)$,
let $\mathbf{A}_{\alpha}$ and $\mathbf{A}_{\beta}$ be the boundary
values of $\mathbf{A}$ obtained by extrapolation from the grains
$\alpha$ and $\beta$, respectively, at the same location. Then $\left[\mathbf{A}\right]\equiv\mathbf{A}_{\alpha}-\mathbf{A}_{\beta}$
represents the jump of $\mathbf{A}$ across the boundary and $\left\langle \mathbf{A}\right\rangle \equiv(\mathbf{A}_{\alpha}+\mathbf{A}_{\beta})/2$
is the average GB value of $\mathbf{A}$. Similar notations are used
for scalar and tensor fields. In particular, diffusion of atoms across
the boundary is described by the phenomenological equation 
\begin{equation}
J_{n}=L_{t}\left[M\right],\label{eq:5}
\end{equation}
while diffusion along the boundary by
\begin{equation}
\mathbf{J}_{b}=-L_{p}\nabla_{b}\left\langle M\right\rangle ,\label{eq:6}
\end{equation}
where $\nabla_{b}\left\langle M\right\rangle $ is the two-dimension
gradient of $\left\langle M\right\rangle $ in the GB plane. In Eqs.(\ref{eq:5})
and (\ref{eq:6}), $L_{t}>0$ and $L_{p}>0$ are the kinetic coefficients
for trans-boundary and intra-boundary diffusion, respectively (note
that they have different dimensions). 

The four diffusion fluxes $\mathbf{J}_{L}^{\alpha}$, $\mathbf{J}_{L}^{\beta}$,
$\mathbf{J}_{b}$ and $\mathbf{J}_{n}$ are subject to two constraints.
First, note that the number of atoms $\nu_{\alpha}$ added to grain
$\alpha$ at the GB (per unit area per unit time) is 
\begin{equation}
\nu_{\alpha}=\dfrac{c_{\alpha}}{\Omega_{\alpha}}\mathbf{n}^{\alpha}\mathbf{\cdot}\left(\mathbf{v}_{b}-\mathbf{v}_{L}^{\alpha}\right)-\mathbf{n}^{\alpha}\mathbf{\cdot}\mathbf{J}_{L}^{\alpha},\label{eq:7}
\end{equation}
while the number of atoms added simultaneously to grain $\beta$ is
\begin{equation}
\nu_{\beta}=-\dfrac{c_{\beta}}{\Omega_{\beta}}\mathbf{n}^{\alpha}\mathbf{\cdot}\left(\mathbf{v}_{b}-\mathbf{v}_{L}^{\beta}\right)+\mathbf{n}^{\alpha}\mathbf{\cdot}\mathbf{J}_{L}^{\beta}.\label{eq:8}
\end{equation}
Here, the dot denotes the inner product (contraction) of vectors or
tensors and $\mathbf{v}_{b}$ is the GB velocity relative to the laboratory
reference frame. The lattice velocities and lattice diffusion fluxes
appearing in these equations are obtained by extrapolation to the
GB from the respective grains. The atoms entering the GB from the
grains can spread along it by GB diffusion. The conservation of atoms
dictates that 
\begin{equation}
\nu_{\alpha}+\nu_{\beta}=-\nabla_{b}\mathbf{\cdot}\mathbf{J}_{b},\label{eq:8a}
\end{equation}
where $\nabla_{b}\mathbf{\cdot}\mathbf{J}_{b}$ is the two-dimensional
divergence of the GB flux $\mathbf{J}_{b}$. In addition, the normal
diffusion flux across the GB can be expressed by \citep{Mishin:2015ad}
\begin{equation}
J_{n}=\dfrac{1}{2}(\nu_{\beta}-\nu_{\alpha}).\label{eq:8b}
\end{equation}
Equations (\ref{eq:8a}) and (\ref{eq:8b}) impose two constraints
on the diffusion fluxes existing at the GB. They can be rewritten
as
\begin{equation}
\left[\dfrac{c}{\Omega}\right]\mathbf{v}_{b}\mathbf{\cdot}\mathbf{n}^{\alpha}-\left[\dfrac{c}{\Omega}\mathbf{v}_{L}\right]\mathbf{\cdot}\mathbf{n}^{\alpha}-\left[\mathbf{J}_{L}\right]\mathbf{\cdot}\mathbf{n}^{\alpha}=-\nabla_{b}\mathbf{\cdot}\mathbf{J}_{b}\label{eq:9}
\end{equation}
and
\begin{equation}
J_{n}=-\left\langle \dfrac{c}{\Omega}\right\rangle \mathbf{v}_{b}\mathbf{\cdot}\mathbf{n}^{\alpha}+\left\langle \dfrac{c}{\Omega}\mathbf{v}_{L}\right\rangle \mathbf{\cdot}\mathbf{n}^{\alpha}+\left\langle \mathbf{J}_{L}\right\rangle \mathbf{\cdot}\mathbf{n}^{\alpha},\label{eq:10}
\end{equation}
respectively.

To formulate the site generation and GB migration equations, we decompose
the lattice velocity jump $\left[\mathbf{v}_{L}\right]$ and the average
lattice velocity $\left\langle \mathbf{v}_{L}\right\rangle $ into
components normal and parallel to the GB: $\left[\mathbf{v}_{L}\right]=\left[\mathbf{v}_{L}\right]_{\bot}+\left[\mathbf{v}_{L}\right]_{||}$
and $\left\langle \mathbf{v}_{L}\right\rangle =\left\langle \mathbf{v}_{L}\right\rangle _{\bot}+\left\langle \mathbf{v}_{L}\right\rangle _{||}$.
Similarly, for the GB traction vector $\mathbf{s}=\mathbf{n}^{\alpha}\mathbf{\cdot}\boldsymbol{\sigma}$
we have $\left\langle \mathbf{s}\right\rangle =\left\langle \mathbf{s}\right\rangle _{\bot}+\left\langle \mathbf{s}\right\rangle _{||}$
and $\left[\mathbf{s}\right]=\left[\mathbf{s}\right]_{\bot}+\left[\mathbf{s}\right]_{||}$.
We will assume that the interiors of the grains and the GB itself
remain in mechanical equilibrium at all times. This is a reasonable
approximation if we focus the attention on 
relatively slow thermally-activated processes such as diffusion. Under mechanical equilibrium
conditions, the traction vector is continuous across the GB plane,
$\left[\mathbf{s}\right]_{\bot}=\left[\mathbf{s}\right]_{||}=\mathbf{0}$,
and thus $\left\langle \mathbf{s}\right\rangle =\mathbf{s}$ \citep{Mishin:2015ad}.
It is assumed that the stress does not cause decohesion of the material,
which would result in the formation of grain boundary cracks, pores
and similar defects.

The normal component of the lattice velocity jump $\left[\mathbf{v}_{L}\right]_{\bot}$
is a measure of the site generation rate at the GB. As shown previously
\citep{Mishin:2015ad}, the thermodynamic driving force for the site
generation at a planar GB is $\left\langle \omega\right\rangle -\mathbf{n}^{\alpha}\mathbf{\cdot}\mathbf{s}_{\bot}$.
The term $\mathbf{n}^{\alpha}\mathbf{\cdot}\mathbf{s}_{\bot}$ captures
the effect of the normal GB stress on the vacancy creation and annihilation
process. The kinetic equation of this process can be written in the
form

\begin{equation}
\mathbf{n}^{\alpha}\mathbf{\cdot}\left[\mathbf{v}_{L}\right]_{\bot}=R\left(\left\langle \omega\right\rangle -\mathbf{n}^{\alpha}\mathbf{\cdot}\mathbf{s}_{\bot}\right),\label{eq:11}
\end{equation}
where $R>0$ is the respective kinetic coefficient.
This coefficient controls the ability of the GB to absorb or generate
vacancies. By contrast to some of the earlier models, we do not assume
that the GB is a ``perfect'' sink/source of vacancies. By varying
the parameter $R$ one can model GBs with a poor sink/source efficiency
(small $R$) and with nearly ideal efficiency (large $R$). Based
on the general knowledge of GBs \citep{Balluffi95} and recent simulations
\citep{Gu:2017aa}, it is expected that the sink/source efficiency
depends on the GB structure, crystallographic parameters, local chemistry
and other factors.

Finally, the GB motion is driven by the grand-canonical potential
jump $\left[\omega\right]$ and the shear stress $\left\langle \mathbf{s}\right\rangle _{||}$
parallel to the GB plane. The kinetic law of GB motion is \citep{Mishin:2015ad}
\begin{equation}
v_{GB}=-L_{GB}\left(\left[\omega\right]+\beta\left|\left\langle \mathbf{s}\right\rangle _{||}\right|\right),\label{eq:12}
\end{equation}
where the kinetic coefficient $L_{GB}>0$ represents the GB mobility
and $v_{GB}=\mathbf{n}^{\alpha}\mathbf{\cdot}\mathbf{v}_{b}-\mathbf{n}^{\alpha}\mathbf{\cdot}\left\langle \mathbf{v}_{L}\right\rangle $
is the GB migration velocity. The latter describes the GB motion relative
to the lattices of the grains and is independent of the choice of
the reference frame. In Eq.(\ref{eq:12}), the second term in the
driving force arises from the shear-coupling effect \citep{Cahn06a,Cahn06b}.
It is assumed that there is a certain direction in the boundary plane,
defined by a unit vector $\mathbf{t}$, such that application of the
shear stress $\left\langle \mathbf{s}\right\rangle _{||}$ parallel
to that direction causes normal GB motion with a speed proportional
to $\left\langle \mathbf{s}\right\rangle _{||}\cdot\mathbf{\mathbf{t}}$.
Conversely, normal GB motion causes relative translation of the grains
with a velocity $\left[\mathbf{v}_{L}\right]_{||}$ parallel to $\mathbf{t}$.
The shear-coupling effect is described by the equation $\left[\mathbf{v}_{L}\right]_{||}\mathbf{\cdot\mathbf{t}}=\beta v_{GB}$,
where the coupling factor $\beta$ depends on the crystallographic
characteristics of the GB \citep{Cahn06a,Cahn06b}. The coupling factor
is not unique to a given GB. It can change with temperature, direction
of the shear stress and other factors as discussed in more detail
in the recent literature \citep{Han:2018aa}.
It can also vary in time as the grain boundary structure changes due
to the absorption or emission of vacancies \citep{Borovikov:2013aa}.
Some GBs respond to applied shear stresses by rigid sliding of the
grains relative to each other. For such GBs, called uncoupled, the
second term in the right-hand side of Eq.(\ref{eq:12}) vanishes.
Instead, the mechanical response of the boundary to applied shear
is described by a sliding law 
\begin{equation}
\left[\mathbf{v}_{L}\right]_{||}=K_{s}\left\langle \mathbf{s}\right\rangle _{||},\label{eq:13}
\end{equation}
where $K_{s}>0$ is the sliding coefficient. The GB migration is then
solely driven by the jump $\left[\omega\right]$. Some GBs that are
coupled at low temperatures can switch from coupling to sliding as
temperature increases \citep{Cahn06b}.

\subsection{Specific model of a bicrystal}

We will next specialize the above equations for a particular case
when the vacancy concentration is small and the elastic deformations
in the grains can be treated in the small-strain approximation \citep{Mishin:2015ad}.
As the reference state of the small-strain tensor $\boldsymbol{\varepsilon}$
we choose the stress-free solid without vacancies ($c=1$). The vacancies
produce an isotropic stress-free deformation
\begin{equation}
\boldsymbol{\varepsilon}^{0}=\dfrac{\Delta\Omega_{v}}{3\Omega^{\prime}}\left(1-c\right)\mathbf{I},\label{eq:14}
\end{equation}
where $\mathbf{I}$ is a $3\times3$ unit tensor, $\Omega^{\prime}$
is the atomic volume in the reference state, and $\Delta\Omega_{v}<0$
is the vacancy relaxation volume under zero stress conditions. The
total lattice strain is 
\begin{equation}
\boldsymbol{\varepsilon}=\boldsymbol{\varepsilon}^{0}+\mathbf{S}:\boldsymbol{\sigma},\label{eq:15}
\end{equation}
where $\mathbf{S}$ is the rank-four tensor of elastic compliances
(the double-dot contraction of two second-rank tensors $\mathbf{a}$
and $\mathbf{b}$ is defined by $\mathbf{a}:\mathbf{b}=\textrm{Tr}(\mathbf{a}\mathbf{b}^{T})$
).

Thermodynamic integration gives the following expressions for the
grand-canonical and diffusion potentials inside the grains \citep{Mishin:2015ad}:
\begin{equation}
\omega=\dfrac{1}{\Omega}\left(kT\ln\dfrac{1-c}{1-c^{*}}-\Delta\Omega_{v}c\sigma_{h}+\dfrac{\Omega^{\prime}}{2}\boldsymbol{\sigma}:\mathbf{S}:\boldsymbol{\sigma}\right),\label{eq:16}
\end{equation}
\begin{equation}
M=M^{*}+kT\ln\dfrac{c(1-c^{*})}{c^{*}(1-c)}+\Delta\Omega_{v}\sigma_{h}.\label{eq:17}
\end{equation}
Here, $c^{*}$ is the equilibrium atomic concentration in stress-free
grains, $M^{*}$ is the respective equilibrium diffusion potential,
$\sigma_{h}=\mathrm{Tr}(\boldsymbol{\sigma})/3$ is the hydrostatic
part of the stress tensor, and 
\begin{equation}
\Omega=\Omega^{\prime}\left[1+\mathrm{Tr}(\boldsymbol{\varepsilon})\right]=\Omega^{\prime}\left[1+\dfrac{\Delta\Omega_{v}}{\Omega^{\prime}}\left(1-c\right)\right]\label{eq:18}
\end{equation}
is the atomic volume. In the latter equation, the term $\mathrm{Tr}(\mathbf{S}:\boldsymbol{\sigma})$
has been neglected due to the small-strain approximation of the elastic
deformation.

We next take into account that the vacancy concentration $c_{v}\equiv(1-c)$
is very small ($c_{v}\ll1$). Accordingly, all equations can be reformulated
in terms of the small parameter $c_{v}$ and simplified. Equations
(\ref{eq:16}) and (\ref{eq:17}) become
\begin{equation}
\omega=\dfrac{kT}{\Omega^{\prime}}\ln\dfrac{c_{v}}{c_{v}^{*}}-\dfrac{\Delta\Omega_{v}}{\Omega^{\prime}}\sigma_{h},\label{eq:19}
\end{equation}
\begin{equation}
M=M^{*}-kT\ln\dfrac{c_{v}}{c_{v}^{*}}+\Delta\Omega_{v}\sigma_{h},\label{eq:20}
\end{equation}
where $c_{v}^{*}=1-c^{*}$ is the equilibrium vacancy concentration
in the absence of stresses. Note that we have neglected the term $\Delta\Omega_{v}c_{v}/\Omega^{\prime}$
in Eq.(\ref{eq:18}) in comparison with unity.

Using the diffusion potential from Eq.(\ref{eq:20}), the lattice
diffusion equations (\ref{eq:3}) and (\ref{eq:4}) take the form
\begin{equation}
\mathbf{J}_{L}=D\dfrac{c_{v}}{\Omega^{\prime}c_{v}^{*}}\nabla c_{v}-D\dfrac{\Delta\Omega_{v}c_{v}^{2}}{\Omega^{\prime}kTc_{v}^{*}}\nabla\sigma_{h},\label{eq:21}
\end{equation}
where $D$ is the diffusion coefficient in stress-free grains with
the equilibrium vacancy concentration.
Equation (\ref{eq:21}) was derived in our previous work \citep{Mishin:2015ad}
by establishing a
relation between the kinetic coefficients $L_{\alpha}$ and $L_{\beta}$
and the diffusion coefficient $D$ by the vacancy mechanism:
\begin{equation}
D=\dfrac{L_{\alpha}kT\Omega c_{v}^{*}}{c_{v}^{2}}\label{eq:21a}
\end{equation}
(and similarly for $L_{\beta}$). The first term in Eq.(\ref{eq:21})
represents the usual concentration-gradient driving force for vacancy
diffusion, whereas the second term captures the stress-gradient effect.
Since we assume that the grains do not contain sinks or sources of
vacancies, the continuity equation
can be applied, leading to the diffusion equation
\begin{eqnarray}
\dfrac{\partial c_{v}}{\partial t} & = & \dfrac{Dc_{v}}{c_{v}^{*}}\nabla^{2}c_{v}+\dfrac{D}{c_{v}^{*}}\left(\nabla c_{v}\right)^{2}-\mathbf{v}_{L}\mathbf{\cdot}\nabla c_{v}\nonumber \\
 & - & \dfrac{D\Delta\Omega_{v}}{kTc_{v}^{*}}c_{v}^{2}\nabla^{2}\sigma_{h}-\dfrac{2D\Delta\Omega_{v}}{kTc_{v}^{*}}c_{v}\nabla\sigma_{h}\cdot\nabla c_{v}.\label{eq:22}
\end{eqnarray}

The site generation and GB migration equations (\ref{eq:11}) and
(\ref{eq:12}) include the discontinuity and the average value of
the grand-canonical potential. To simplify the calculations, we will
assume that the hydrostatic stress is continuous across the boundary.
This assumption is more strict than the mechanical equilibrium conditions
alone, which only requires that the traction vector be continuous.\footnote{A discontinuity of the hydrostatic stress would create an additional
driving force for grain boundary migration. This force is quadratic
in stress and under real conditions is usually small.} Under this assumption, 
\begin{equation}
\left[\omega\right]=\dfrac{kT}{2\Omega^{\prime}}\ln\dfrac{c_{v}^{\alpha}}{c_{v}^{\beta}},\label{eq:25}
\end{equation}
\begin{equation}
\left\langle \omega\right\rangle =\dfrac{kT}{2\Omega^{\prime}}\ln\dfrac{c_{v}^{\alpha}c_{v}^{\beta}}{\left(c_{v}^{*}\right)^{2}}-\dfrac{\Delta\Omega_{v}\sigma_{h}}{\Omega^{\prime}}.\label{eq:26}
\end{equation}
In these equations, $c_{v}^{\alpha}$ and $c_{v}^{\beta}$ are the
GB values of the vacancy concentration extrapolated from the grains.
Similarly, $\left[M\right]$ and $\left\langle M\right\rangle $ appearing
in the trans-boundary and intra-boundary diffusion equations are given
by
\begin{equation}
\left[M\right]=-\dfrac{kT}{2}\ln\dfrac{c_{v}^{\alpha}}{c_{v}^{\beta}},\label{eq:27}
\end{equation}
\begin{equation}
\left\langle M\right\rangle =M^{*}-\dfrac{kT}{2}\ln\dfrac{c_{v}^{\alpha}c_{v}^{\beta}}{\left(c_{v}^{*}\right)^{2}}+\Delta\Omega_{v}\sigma_{h}.\label{eq:28}
\end{equation}

\section{Numerical study\label{sec:Numerical-study}}

\subsection{The governing equations}

\begin{figure}
\noindent \begin{centering}
\includegraphics[scale=0.65]{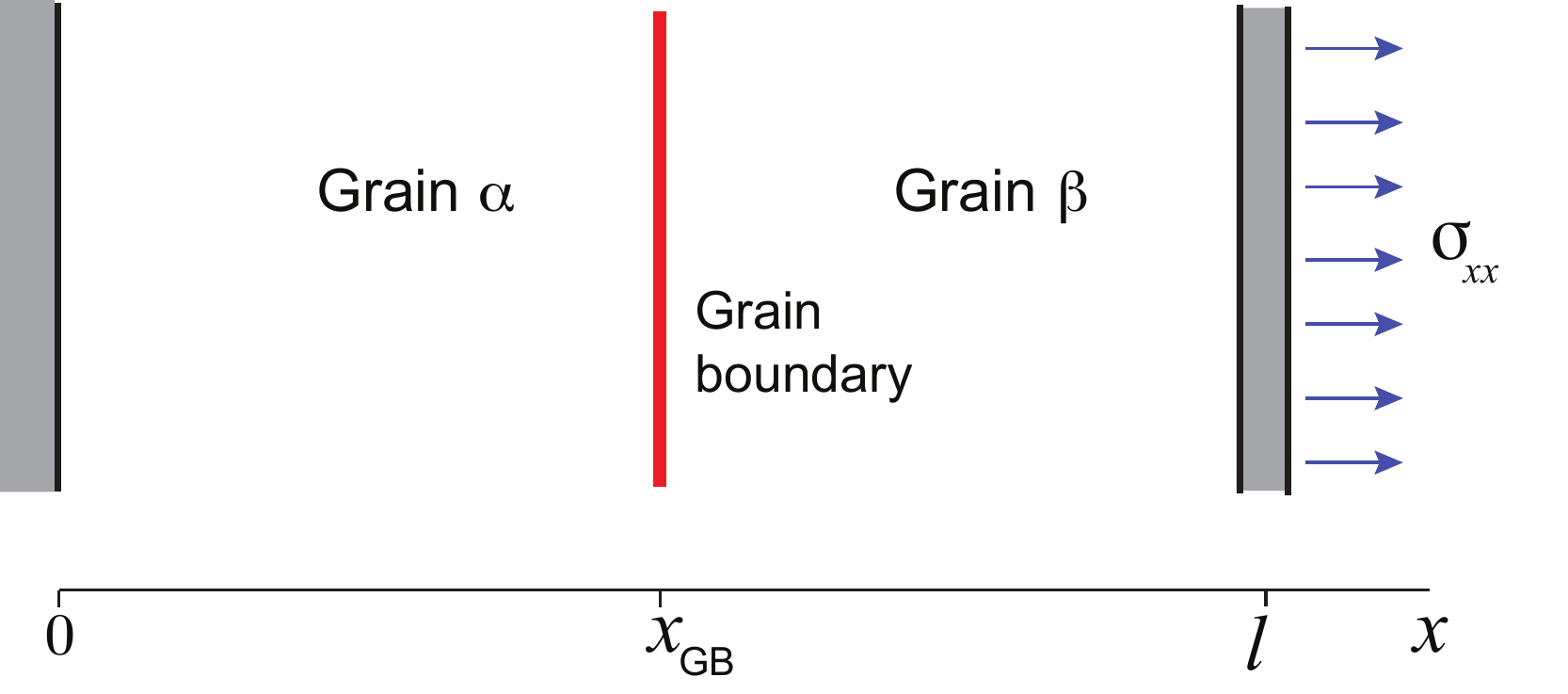}
\par\end{centering}
\caption{Schematic bicrystal with a plane grain boundary. The left bar indicates
the fixed wall. The system is subject to uniform uniaxial stress $\sigma_{xx}$
in the $x$-direction.\label{fig:bicrystal}}
\end{figure}

The following case is chosen for a parametric numerical study. Suppose
the grains $\alpha$ and $\beta$ are separated by a plane incoherent
GB normal to the Cartesian direction $x$ (Fig.~\ref{fig:bicrystal}).
At the left end, the grain $\alpha$ is attached to a fixed wall at
$x=0$. The reference frame is also attached to the wall. At the right
end, the grain $\beta$ terminates at a movable wall (piston) capable
of exerting a prescribed normal stress $\sigma_{11}$ without shear
($\sigma_{12}=\sigma_{13}=0$). Let $x_{GB}(t)$ denote the current
GB position and $x=l(t)$ be the right end of grain $\beta$. The
lateral dimensions of the grains are assumed to be much larger than
$l$. All intensive properties of the system depend only on the coordinate
$x$, making the problem effectively one-dimensional. The temperature
is fixed.

Suppose the lateral dimensions of the system were initially adjusted
so that in the absence of the external force ($\sigma_{11}=0$), the
system reached thermodynamic equilibrium with stress-free grains.
The equilibrium vacancy concentration in this state is $c_{v}^{*}$.
The lateral dimensions of the grains were then fixed once and for
all. Next, the vacancy concentration was modified (e.g., by irradiation)
and/or a stress $\sigma_{11}$ was applied. This creates a new initial
state whose subsequent evolution we wish to investigate. Note that
during this evolution, the boundary conditions ensure that the lateral
lattice strains, $\varepsilon_{22}$ and $\varepsilon_{33}$, remain
fixed at the initial, stress-free value {[}cf.~Eq.(\ref{eq:14}){]},
\begin{equation}
\varepsilon_{22}=\varepsilon_{33}=\dfrac{\Delta\Omega_{v}}{3\Omega^{\prime}}c_{v}^{*}.\label{eq:29}
\end{equation}

To simplify the calculations, the grains will be treated as homogeneous,
elastically isotropic media with a Young modulus $E$ and Poisson's
factor $\nu$. As before, it is assumed that mechanical equilibrium
is always maintained ($\nabla\cdot\boldsymbol{\sigma}=\mathbf{0}$).
Consequently, $\sigma_{22}$ and $\sigma_{33}$ must be equal functions
of only $x$ ($\sigma_{22}(x)=\sigma_{33}(x)$), $\sigma_{11}$ must
be uniform throughout the system and equal to its boundary value at
$x=l$, and all other stress components must be zero. Furthermore,
as was done above, we will neglect the stress-free strain produced
by the vacancies. In this approximation $\varepsilon_{22}=\varepsilon_{33}=0$
and the isotropic Hooke's law is readily solved for the normal strain
$\varepsilon_{11}$ and lateral stresses $\sigma_{22}$ and $\sigma_{33}$:

\begin{equation}
\varepsilon_{11}=\dfrac{(1+\nu)(1-2\nu)}{E(1-\nu)}\sigma_{11},\label{eq:30}
\end{equation}
\begin{equation}
\sigma_{22}=\sigma_{33}=\dfrac{\nu}{1-\nu}\sigma_{11}.\label{eq:31}
\end{equation}
Thus, all stresses and strains in the system are uniform, time-independent,
and are uniquely defined by the applied normal stress $\sigma_{11}$.
The hydrostatic part of the stress tensor is 
\begin{equation}
\sigma_{h}=\dfrac{1}{3}\left(\dfrac{1+\nu}{1-\nu}\right)\sigma_{11}.\label{eq:32}
\end{equation}

The lattice velocity field $v_{L}(x,t)$ is generally related to the
deformation rate by $\partial v_{L}/\partial x=\dot{\varepsilon}_{11}$.
However, due to the neglect of the stress-free strain, $\varepsilon_{11}$
is constant and thus $\partial v_{L}/\partial x=0$. Each grain moves
with a uniform velocity as a rigid body. Because grain $\alpha$ is
attached to the wall, its velocity vanishes, while grain $\beta$
moves with a time-dependent velocity $v_{L}^{\beta}(t)$\@. 

Diffusion inside the grains is described by Eqs.(\ref{eq:21}) and
(\ref{eq:22}), which simplify to
\begin{equation}
J_{L}=\dfrac{Dc_{v}}{\Omega^{\prime}c_{v}^{*}}\dfrac{\partial c_{v}}{\partial x}\label{eq:34}
\end{equation}
and
\begin{equation}
\dfrac{\partial c_{v}}{\partial t}=\dfrac{Dc_{v}}{c_{v}^{*}}\dfrac{\partial^{2}c_{v}}{\partial x^{2}}+\dfrac{D}{c_{v}^{*}}\left(\dfrac{\partial c_{v}}{\partial x}\right)^{2}-v_{L}\dfrac{\partial c_{v}}{\partial x},\label{eq:33}
\end{equation}
respectively. Note that the stress derivatives appearing in Eqs.(\ref{eq:21})
and (\ref{eq:22}) have vanished due to the spatial uniformity of
the stress tensor. 

Turning to GB processes, the axial symmetry of the problem precludes
lateral diffusion in the GB. Thus $\mathbf{J}_{b}=\mathbf{0}$ and
the mass conservation equation (\ref{eq:9}) takes the form
\begin{equation}
\left(c_{v}^{\beta}-c_{v}^{\alpha}\right)\dfrac{dx_{GB}}{dt}+v_{L}^{\beta}+\dfrac{D}{c_{v}^{*}}\left(c_{v}^{\beta}\left(\dfrac{\partial c_{v}}{\partial x}\right)_{\beta}-c_{v}^{\alpha}\left(\dfrac{\partial c_{v}}{\partial x}\right)_{\alpha}\right)=0.\label{eq:35}
\end{equation}
Note that in the second term we replaced $c_{\beta}v_{L}^{\beta}$
by $v_{L}^{\beta}$ due to the smallness of the vacancy concentration.
The normal flux of atoms across the boundary is 
\begin{equation}
J_{n}=-\dfrac{1}{\Omega^{\prime}}\dfrac{dx_{GB}}{dt}+\dfrac{v_{L}^{\beta}}{2\Omega^{\prime}}+\dfrac{D}{2c_{v}^{*}\Omega^{\prime}}\left(c_{v}^{\alpha}\left(\dfrac{\partial c_{v}}{\partial x}\right)_{\alpha}+c_{v}^{\beta}\left(\dfrac{\partial c_{v}}{\partial x}\right)_{\beta}\right),\label{eq:36}
\end{equation}
where we have used Eq.(\ref{eq:10}) and approximated $c_{\alpha}+c_{\beta}\approx2$
and $c_{\beta}\approx1$. Using $\left[M\right]$ from Eq.(\ref{eq:27}),
the kinetic equation of trans-boundary diffusion (\ref{eq:5}) becomes
\begin{equation}
-2\dfrac{dx_{GB}}{dt}+v_{L}^{\beta}+\dfrac{D}{c_{v}^{*}}\left(c_{v}^{\alpha}\left(\dfrac{\partial c_{v}}{\partial x}\right)_{\alpha}+c_{v}^{\beta}\left(\dfrac{\partial c_{v}}{\partial x}\right)_{\beta}\right)=-L_{t}^{\prime}\ln\dfrac{c_{v}^{\alpha}}{c_{v}^{\beta}}\label{eq:38}
\end{equation}
with the kinetic coefficient  $L_{t}^{\prime}=kT\Omega^{\prime}L_{t}$.
The vacancy generation equation (\ref{eq:11}) simplifies to 
\begin{equation}
v_{L}^{\beta}=-R\left(\left\langle \omega\right\rangle -\sigma_{11}\right),\label{eq:39}
\end{equation}
while the GB migration equation (\ref{eq:12}) reduces to 
\begin{equation}
v_{GB}=\dfrac{dx_{GB}}{dt}-\dfrac{1}{2}v_{L}^{\beta}=-L_{GB}\left[\omega\right].\label{eq:40}
\end{equation}
Equations (\ref{eq:39}) and (\ref{eq:40}) show that, under vacancy
equilibrium conditions, $\omega$ must be uniform throughout the system
and equal to $\sigma_{11}$. The equilibrium vacancy concentration
$c_{v}^{eq}$ must be also uniform and satisfy the equation
\begin{equation}
\dfrac{kT}{\Omega^{\prime}}\ln\dfrac{c_{v}^{eq}}{c_{v}^{*}}=\dfrac{\Delta\Omega_{v}}{\Omega^{\prime}}\sigma_{h}+\sigma_{11},\label{eq:41}
\end{equation}
where we used Eq.(\ref{eq:26}) for $\left\langle \omega\right\rangle $.
When $\sigma_{11}=0$, $c_{v}^{eq}$ reduces to its stress-free value
$c_{v}^{*}$.

If the term with $\Delta\Omega_{v}$ could be dropped, Eq.(\ref{eq:41})
would reduce to Herring's relation for the effect of applied stress
on the vacancy concentration in solids \citep{herring49,Herring1950}.
However, this term is generally non-negligible since $\Delta\Omega_{v}$
is usually not much smaller than $\Omega^{\prime}$. This term captures
the additional effect of lateral elastic deformation on the equilibrium
vacancy concentration in stressed solids. In the bicrystal considered
here, the lateral stresses arise due to the constraint opposing the
Poisson deformation caused by the normal stress. We note in passing
that in other cases lateral stresses can be more significant. In epitaxial
thin films, lateral (in-plane) stresses can be high while $\sigma_{11}$
is zero, or at least much smaller. Under such conditions, it is the
lateral stress that dictates the vacancy concentration in the film.

In the final form, the vacancy generation and GB migration equations
take the forms
\begin{equation}
v_{L}^{\beta}=-\dfrac{RkT}{2\Omega^{\prime}}\ln\dfrac{c_{v}^{\alpha}c_{v}^{\beta}}{\left(c_{v}^{eq}\right)^{2}}\label{eq:42}
\end{equation}
and
\begin{equation}
\dfrac{dx_{GB}}{dt}-\dfrac{1}{2}v_{L}^{\beta}=-\dfrac{kTL_{GB}}{2\Omega^{\prime}}\ln\dfrac{c_{v}^{\alpha}}{c_{v}^{\beta}},\label{eq:43}
\end{equation}
respectively.

\subsection{The numerical method for solving the equations}

The kinetic equations formulated above form a closed system that can
be solved numerically as follows. The diffusion equation (\ref{eq:33})
is solved inside each grain individually, with $v_{L}^{\alpha}\equiv0$
in grain $\alpha$ and the instantaneous lattice velocity $v_{L}^{\beta}(t)$
in grain $\beta$. Note that both equations must be solved in domains
with moving boundaries. At $x=0$ and $x=l(t)$, the zero-flux conditions
$\partial c_{v}/\partial x\equiv0$ are applied. We need two more
boundary conditions at the GB and must know the velocities $v_{L}^{\beta}$
and $\dot{x}_{GB}$. This information is provided by the four equations
(\ref{eq:35}), (\ref{eq:38}), (\ref{eq:42}) and (\ref{eq:43}).
The initial condition for the vacancy concentration can be chosen
arbitrarily. We thus
need to solve a system of partial differential equations in two variable-size
domains simultaneously with differential equations describing the
boundary conditions at the moving interfaces. Similar equations often
arise when describing the dynamics of solidification, phase precipitation,
and other processes involving a phase formation. In the present case,
both grains represent the same phase but have variable dimensions
due to the grain boundary motion and the lattice site formation and
annihilation at the boundary.

Numerical simulations were performed using an explicit-in-time finite
difference approximation to the differential equations, coupled to
a nonlinear solver that treats the interface conditions implicitly.
More specifically, the spatial domains $0<x<x_{GB}(t)$ and $x_{GB}(t)<x<l(t)$
were mapped to fixed computational domains $0<z<1$ and $1<z<2$,
respectively, via the transformations
\begin{equation}
z=\dfrac{x}{x_{GB}(t)},\qquad x<x_{GB}(t),\label{eq:z1}
\end{equation}
\begin{equation}
z=1+\dfrac{x-x_{GB}(t)}{l(t)-x_{GB}(t)},\qquad l(t)>x>x_{GB}(t),\label{eq:z2}
\end{equation}
which result in a new set of differential equations that explicitly
involve coefficients that depend on $x_{GB}(t)$ and $l(t)$. Given
the solution at time $t=n\Delta t$, the vacancy concentration field
at time $t=(n+1)\Delta t$ was computed at interior points of the
grains (excluding the GB) by using a first-order-accurate explicit
time differencing scheme (``forward Euler method''), together with
no-flux boundary conditions at the sample ends. To complete the solution
at this time step, the four additional unknowns $c_{v}^{\alpha}$,
$c_{v}^{\beta}$, $v_{L}^{\beta}$ and $dx_{GB}/dt$ were solved by
applying Newton's method to the four equations (\ref{eq:35}), (\ref{eq:38}),
(\ref{eq:42}) and (\ref{eq:43}). The 
derivatives $\left(\partial c_{v}/\partial x\right)_{\alpha}$ and
$\left(\partial c_{v}/\partial x\right)_{\beta}$ appearing in Eqs.(\ref{eq:35})
and (\ref{eq:38})
are evaluated using one-sided derivatives using the latest interior values
of the vacancy concentration. Good initial guesses for the nonlinear
solver are available from the previous time step. Given the updated
velocities of the lattice and the GB, new values of $l(t)$ and $x_{GB}(t)$
were obtained by again using a first-order time difference to complete
the solution at the new time level. Numerical stability considerations
restrict the size of the time step $\Delta t$ to scale with the square
of the spatial mesh, but this is not a serious limitation in one spatial
dimension. For numerical convenience, in all calculations reported
below, the equilibrium vacancy concentration is $c_{v}^{*}=10^{-4}$.
For comparison, in most materials the vacancy concentration near the
melting point ranges from $10^{-3}$ to $10^{-5}$ \citep{Kraftmakher:1998aa}.

\subsection{Dimensionless form of the governing equations}

The numerical results will be analyzed in terms of the following dimensionless
variables: time $\tau=Dt/l_{0}^{2}$, coordinate $\xi=x/l_{0}$, the
system length $\xi_{l}=l/l_{0}$, the GB position $\xi_{GB}=x_{GB}/l_{0}$,
the lattice velocity $\eta=v_{L}^{\beta}l_{0}/D$, the stress $s=\sigma_{11}\Omega^{\prime}/kT$,
and the normalized vacancy concentration $C=c_{v}/c_{v}^{*}$, where
$l_{0}$ is the initial length of the bicrystal. The dimensionless
GB velocity is 
\begin{equation}
\eta_{GB}=\dfrac{v_{GB}l_{0}}{D}=\dfrac{d\xi_{GB}}{d\tau}-\dfrac{\eta}{2}.\label{eq:43a}
\end{equation}

The dimensionless diffusion equations of the model become
\begin{equation}
\dfrac{\partial C}{\partial\tau}=C\dfrac{\partial^{2}C}{\partial\xi^{2}}+\left(\dfrac{\partial C}{\partial\xi}\right)^{2},\:\:\:\mathrm{grain\:\alpha},\label{eq:165}
\end{equation}
\begin{equation}
\dfrac{\partial C}{\partial\tau}=C\dfrac{\partial^{2}C}{\partial\xi^{2}}+\left(\dfrac{\partial C}{\partial\xi}\right)^{2}-\eta\dfrac{\partial C}{\partial\xi},\:\:\:\mathrm{grain\:\beta}.\label{eq:166}
\end{equation}
The vacancy concentration fields in the grains are subject to the
boundary conditions $\partial C/\partial\xi\equiv0$ at $\xi=0$ and
$\xi=\xi_{l}(t)$, and to the following boundary equations at $\xi=\xi_{GB}(t)$:
\begin{equation}
-2\dfrac{d\xi_{GB}}{d\tau}+\eta+c_{v}^{*}\left(C_{\alpha}\left(\dfrac{\partial C}{\partial\xi}\right)_{\alpha}+C_{\beta}\left(\dfrac{\partial C}{\partial\xi}\right)_{\beta}\right)=-\lambda_{t}\ln\dfrac{C_{\alpha}}{C_{\beta}},\label{eq:172}
\end{equation}
\begin{equation}
c_{v}^{*}\left(C_{\beta}-C_{\alpha}\right)\dfrac{d\xi_{GB}}{d\tau}+\eta+c_{v}^{*}\left(C_{\beta}\left(\dfrac{\partial C}{\partial\xi}\right)_{\beta}-C_{\alpha}\left(\dfrac{\partial C}{\partial\xi}\right)_{\alpha}\right)=0,\label{eq:173}
\end{equation}
\begin{equation}
\eta=-\rho\left(\dfrac{1}{2}\ln C_{\alpha}C_{\beta}-\dfrac{b}{3}\left(\dfrac{1+\nu}{1-\nu}\right)s-s\right),\label{eq:167}
\end{equation}
\begin{equation}
\dfrac{d\xi_{GB}}{d\tau}=\dfrac{1}{2}\eta-\lambda\ln\dfrac{C_{\alpha}}{C_{\beta}}.\label{eq:168}
\end{equation}
These equations depend on the dimensionless parameters $\lambda_{t}=(l_{0}/D)L_{t}^{\prime}$,
$\lambda=(kTl_{0}/D\Omega^{\prime})L_{GB}$, $\rho=(kTl_{0}/D\Omega^{\prime})R$
and $b=\Delta\Omega_{v}/\Omega^{\prime}$. These parameters control
the trans-boundary diffusion, the GB mobility, the rate of vacancy
generation/annihilation at the GB, and the vacancy relaxation volume,
respectively.

Note that $\xi_{l}(0)=1$ by definition. At any moment of time, the
current length of the sample can be found by the integration
\begin{equation}
\xi_{l}(\tau)=1+\intop_{0}^{\tau}\eta(\tau)d\tau.\label{eq:174}
\end{equation}

The evolution of the system depends on the initial vacancy distribution
$C(\xi,0)$ in each grain,
the initial GB position $\xi_{GB}(0)$, and the applied stress $s$.
The system eventually reaches equilibrium with a stress-dependent
uniform vacancy concentration $c_{v}^{eq}$. The latter can be obtained
from Eq.(\ref{eq:167}) by setting $\eta=0$, which gives
\begin{equation}
\ln C^{eq}\equiv\ln\dfrac{c_{v}^{eq}}{c_{v}^{*}}=\left(1+\dfrac{b}{3}\left(\dfrac{1+\nu}{1-\nu}\right)\right)s.\label{eq:171}
\end{equation}
Without loss of generality, we will assume that $s=0$ and thus $C^{eq}=1$.
Under this condition, the dynamics are only driven by diffusive redistribution
of the vacancies between the grains and absorption of non-equilibrium
vacancies by the grain boundary. For nonzero stress, the solution
can be readily obtained by redefining the dimensionless vacancy concentration
as $C=c_{v}/c_{v}^{eq}$, where $c_{v}^{eq}$ refers to the equilibrium
state under $s\neq0$. The evolution equations are then obtained by
simply replacing $c_{v}^{*}$ by $c_{v}^{eq}$ and re-interpreting
$D$ as the diffusion coefficient in the stressed lattice with the
vacancy concentration $c_{v}^{eq}$.

\subsection{The normal modes of relaxation}

The linear stability analysis of the model presented in Appendix B
shows that there are two non-trivial relaxation modes near the system
equilibrium. The linear stability analysis assumes that the GB and
lattice velocities and the deviation of the vacancy concentration
from equilibrium all decay with time in proportion to $\exp(-\kappa^{2}\tau)$.
In the so-called $\kappa_{\rho}$ mode, the decaying solutions are
(cf.~Eq.(\ref{eq:Lin_Syst_1-2}))
\begin{equation}
C_{\alpha}=1+\hat{C}\exp(-\kappa_{\rho}^{2}\tau)\cos\xi\kappa_{\rho},\label{eq:profile_rho_1}
\end{equation}
\begin{equation}
C_{\beta}=1+\hat{C}\exp(-\kappa_{\rho}^{2}\tau)\cos(\xi-1)\kappa_{\rho},\label{eq:profile_rho_2}
\end{equation}
\begin{equation}
\xi_{GB}=\dfrac{1}{2}+\dfrac{\rho\hat{C}}{2\kappa_{\rho}^{2}}\exp(-\kappa_{\rho}^{2}\tau)\cos\dfrac{\kappa_{\rho}}{2},\label{eq:GB_vel_1}
\end{equation}
\begin{equation}
\eta=-\rho\hat{C}\exp(-\kappa_{\rho}^{2}\tau)\cos\dfrac{\kappa_{\rho}}{2}.\label{eq:lattice_vel_1}
\end{equation}
Here, $\hat{C}$ is the initial excess of the vacancy concentration
over the equilibrium value at the ends of the sample. The boundary
is initially at $\xi=1/2$, where the vacancy excess is $\hat{C}\cos\kappa_{\rho}/2$.
The wavenumber $\kappa_{\rho}$ is the root of the transcendental
equation 
\begin{equation}
\dfrac{2Z_{\rho}}{\kappa_{\rho}}=\tan\dfrac{\kappa_{\rho}}{2},\label{eq:Trans_rho}
\end{equation}
with the constant parameter
\begin{equation}
Z_{\rho}=\dfrac{\rho}{4c_{v}^{*}}.\label{eq:Z_rho_def}
\end{equation}
In this mode, the vacancy concentration is a continuous (no interface
gap) and even function about the GB position. The dominant relaxation
process is the vacancy generation or absorption by the GB. The boundary
itself does not move relative to the mean velocity of the grains ($\eta_{GB}=d\xi_{GB}/d\tau-\eta/2=0$).

The second normal mode is referred to as the $\kappa_{\lambda}$ mode,
in which (cf.~Eq.(\ref{eq:Lin_Syst_1-2-1}))
\begin{equation}
C_{\alpha}=1+\hat{C}\exp(-\kappa_{\lambda}^{2}\tau)\cos\xi\kappa_{\lambda},\label{eq:Profile_lambda_1}
\end{equation}
\begin{equation}
C_{\beta}=1-\hat{C}\exp(-\kappa_{\lambda}^{2}\tau)\cos(\xi-1)\kappa_{\lambda},\label{eq:profile_lambda_2}
\end{equation}
\begin{equation}
\xi_{GB}=\dfrac{1}{2}+\dfrac{2\lambda\hat{C}}{\kappa_{\lambda}^{2}}\exp(-\kappa_{\lambda}^{2}\tau)\cos\dfrac{\kappa_{\lambda}}{2},\label{eq:GB_vel_2}
\end{equation}
\begin{equation}
\eta=0,\label{eq:lattice_vel_2}
\end{equation}
where $\kappa_{\lambda}$ is the root of the equation 
\begin{equation}
\dfrac{2Z_{\lambda}}{\kappa_{\lambda}}=\tan\dfrac{\kappa_{\lambda}}{2},\label{eq:Trans_lambda}
\end{equation}
with 
\begin{equation}
Z_{\lambda}=\dfrac{2\lambda+\lambda_{t}}{2c_{v}^{*}}.\label{eq:Z_lambda_def}
\end{equation}
There is no vacancy generation or absorption at the boundary. The
relaxation occurs by vacancy diffusion in the grains and across the
GB. The initial vacancy concentration is $1+\hat{C}$ at $\xi=0$
and $1-\hat{C}$ at $\xi=1$. Thus, the vacancies are oversaturated
at one end of the sample and undersaturated at the other. The vacancy
concentration profile is an odd function relative to the GB position,
with the concentration
gap
\[
\Delta C=C_{\beta}-C_{\alpha}=\hat{C}\exp(-\kappa_{\lambda}^{2}\tau)\cos\dfrac{\kappa_{\lambda}}{2}
\]
at the boundary. During the relaxation process, the
concentration gap narrows down and eventually vanishes
as the vacancy concentration reaches the equilibrium value $C_{\beta}=C_{\alpha}=1$.
Note that this process is accompanied by GB migration with the velocity
\[
\eta_{GB}=\dfrac{d\xi_{GB}}{d\tau}=-2\lambda\Delta C
\]
driven by the vacancy concentration gap
$\Delta C$.

The present model contains three dimensionless kinetic coefficients:
$\lambda$, $\lambda_{t}$ and $\rho$, which control the GB mobility,
trans-boundary diffusion and vacancy generation/adsorption, respectively.
The linear stability analysis identifies two combinations of these
parameters, $Z_{\rho}$ and $Z_{\lambda}$, which can be the most
effective predictors of the system evolution. Therefore, the numerical
cases presented below were primarily chosen based on the values of
$Z_{\rho}$ and $Z_{\lambda}$.

\subsection{Numerical results}

\subsubsection{Relaxation modes}

\begin{figure}
\noindent \begin{centering}
\textbf{(a)} \enskip{}\includegraphics[width=0.6\textwidth]{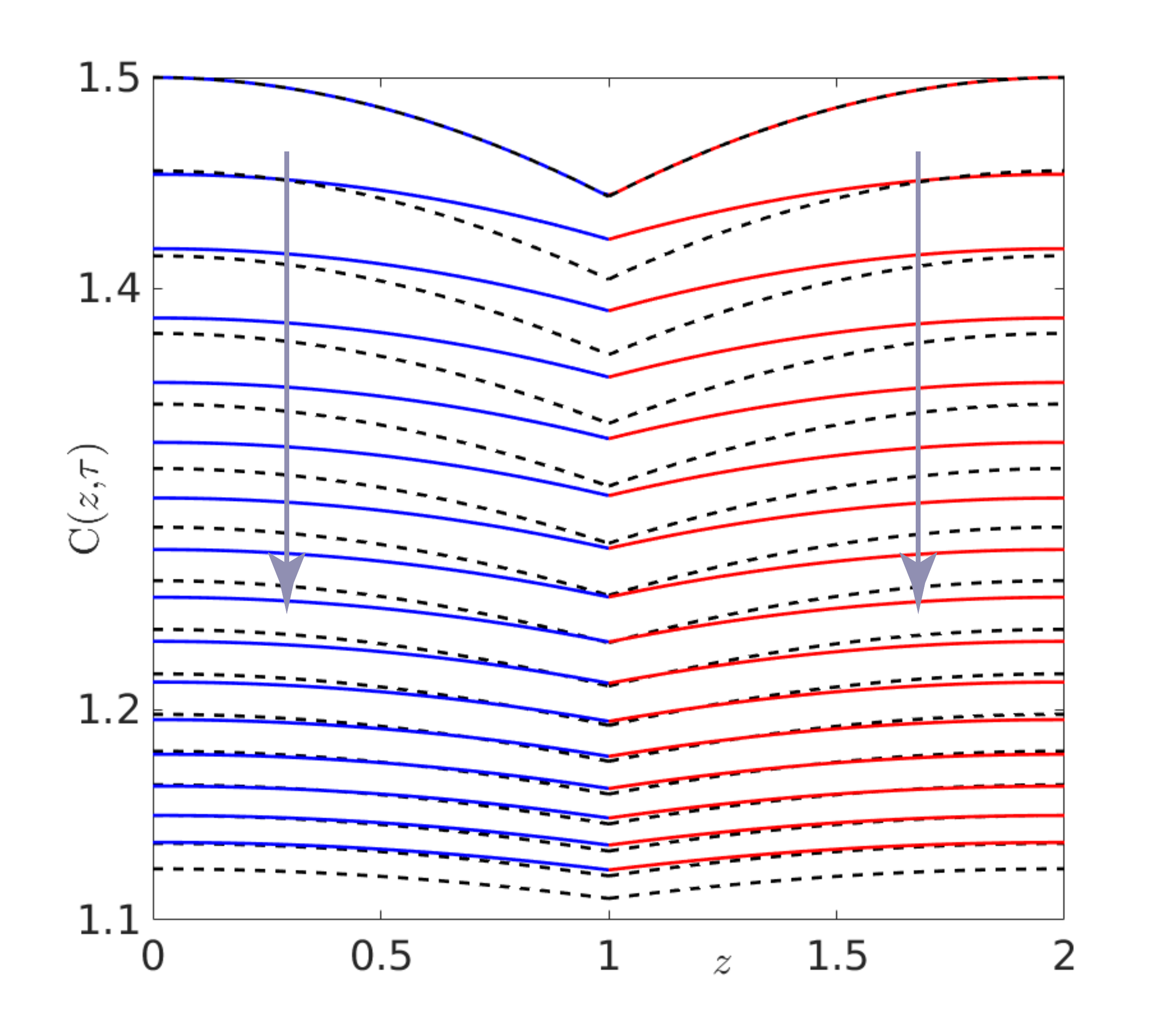}
\par\end{centering}
\bigskip{}

\noindent \begin{centering}
\textbf{(b)} \enskip{}\includegraphics[width=0.7\textwidth]{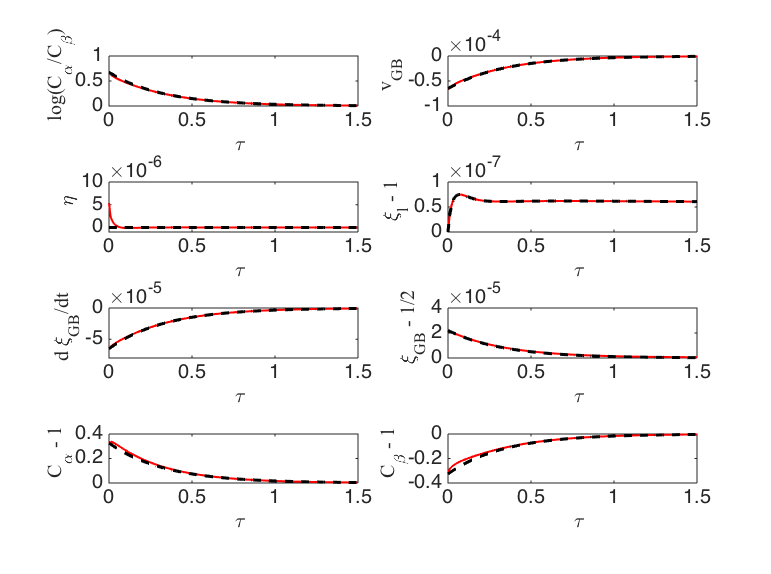}
\par\end{centering}
\caption{(a) Comparison of the $\kappa_{\rho}$ normal mode solution (dashed
black curves) and the nonlinear numerical solution (solid blue and
red curves) for the vacancy concentration fields in each grain. The
concentration profiles are shown at times $\tau$ ranging from 0 to
1.5 at 0.1 increments.
The arrows indicate the direction of time. The model parameters are
$\lambda=\lambda_{t}=0$ and $\rho=10^{-4}$ (thus $Z_{\rho}=1/4$
and $Z_{\lambda}=0$) and the initial perturbation has the amplitude
$\hat{C}=0.5$. (b) Comparison of the same normal mode solution (dashed
red curves) and nonlinear numerical solution (solid black curves)
for various interfacial quantities.\label{fig:k_rho-mode}}

\end{figure}

We will start by illustrating
the two normal modes mentioned above. To create the $\kappa_{\rho}$
mode, we choose $\lambda=\lambda_{t}=0$ and $\rho=10^{-4}$. For
these values of the parameters, we have $Z_{\rho}=1/4$ and $Z_{\lambda}=0$.
The initial distribution is given by Eqs.(\ref{eq:profile_rho_1})
and (\ref{eq:profile_rho_2}) with $\tau=0$ and $\hat{C}=0.5$, so
the maximum vacancy concentration in the system is $1.5c_{v}^{*}$.
The GB velocity is initialized by $d\xi_{GB}/d\tau=0$. The vacancy
concentration profiles obtained by numerical solution of Eqs.(\ref{eq:165})-(\ref{eq:168})
are plotted as a function of time in Fig.~\ref{fig:k_rho-mode}.
For comparison, the dashed lines show predictions of the normal mode
equations (\ref{eq:profile_rho_1}) and (\ref{eq:profile_rho_2}).
Since the grain sizes can vary in time, it is convenient to map the
vacancy concentration profiles in the grains $\alpha$ and $\beta$
onto fixed-size intervals $[0,1]$ and $[1,2]$, respectively, by
the coordinate transformations (cf.~Eqs.(\ref{eq:z1}) and (\ref{eq:z2}))
\begin{equation}
z=\dfrac{\xi}{\xi_{GB}(\tau)},\qquad\xi<\xi_{GB}(\tau),\label{eq:map1}
\end{equation}
\begin{equation}
z=1+\dfrac{\xi-\xi_{GB}(\tau)}{\xi_{l}(\tau)-\xi_{GB}(\tau)},\qquad\xi_{l}(\tau)>\xi>\xi_{GB}(\tau).\label{eq:map2}
\end{equation}
This transformation was applied to all vacancy concentration plots
shown in the paper. The functions $\xi_{GB}(\tau)$ and $\xi_{l}(\tau)$
are then plotted in separate panels.

The plots in Fig.~\ref{fig:k_rho-mode}(a) demonstrate the formation
of a minimum in the vacancy concentration located at the GB, which
approaches the equilibrium value causing the profile to become deeper
and wider with time until the entire system reaches the vacancy equilibrium.
The vacancy absorption at the boundary results in elimination of atomic
layers on either side of the GB plane and thus shrinkage of both grains
by equal amount, as illustrated in Fig.~\ref{fig:k_rho-mode}(b).
The GB migration velocity relative to the grains remains zero within
the numerical accuracy. Note the reasonable agreement between the
numerical solution of the nonlinear model and the normal mode predictions.
The differences between the two solutions arise from the fact that
the normal mode only represents the asymptotic behavior in the limit
of small deviations from equilibrium. Much closer agreement was found
in a similar comparison for a small perturbation by $\hat{C}=0.5\times10^{-4}$
(see Fig.~1 in the Supplementary Material \citep{Supplementary-Material}).

\begin{figure}
\noindent \begin{centering}
\textbf{(a)} \enskip{}\includegraphics[width=0.6\textwidth]{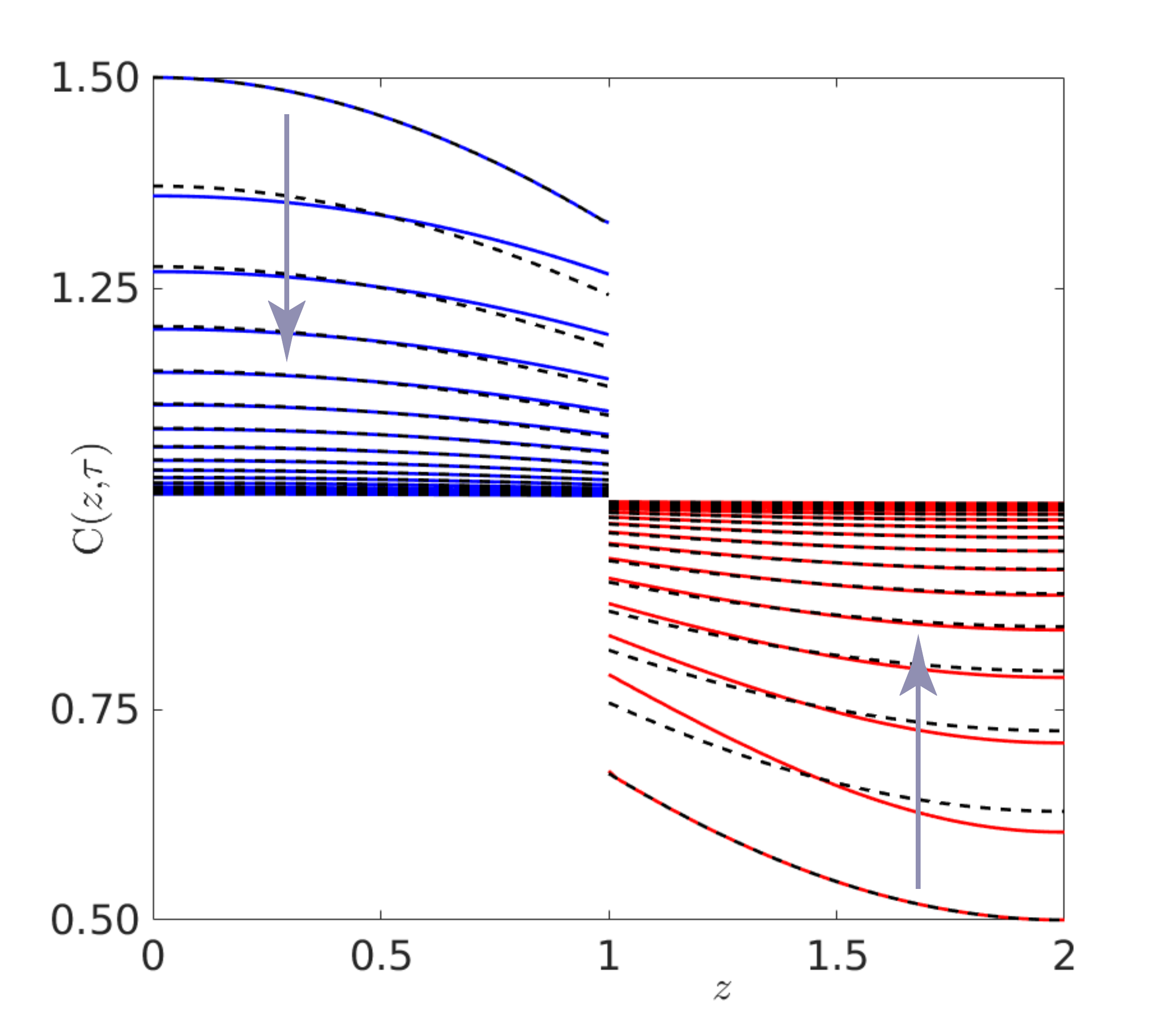}
\par\end{centering}
\bigskip{}

\noindent \begin{centering}
\textbf{(b)} \enskip{}\includegraphics[width=0.7\textwidth]{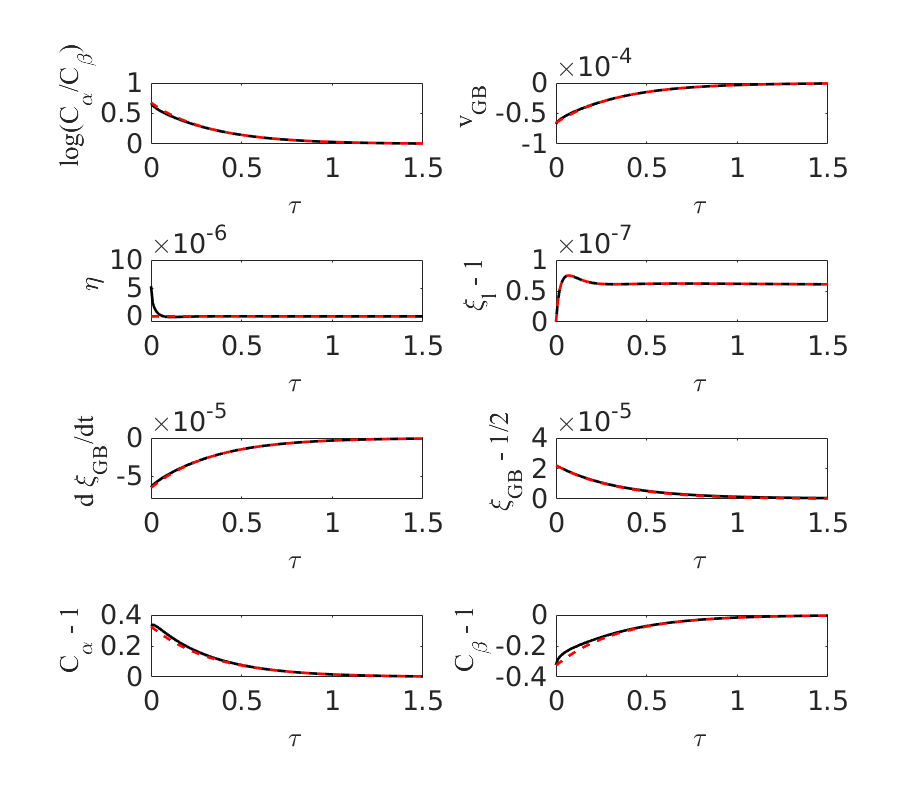}
\par\end{centering}
\caption{(a) Comparison of the $\kappa_{\lambda}$ normal mode solution (dashed
black curves) and the nonlinear numerical solution (solid blue and
red curves) for the vacancy concentration fields in each grain. The
concentration profiles are shown at times $\tau$ ranging from 0 to
1.5 at 0.1 increments.
The arrows indicate the direction of time. The model parameters are
$\lambda=10^{-4}$, $\lambda_{t}=0$ and $\rho=10^{-4}$ (thus $Z_{\rho}=1/4$
and $Z_{\lambda}=1$). (b) Comparison of the same normal mode solution
(dashed red curves) and nonlinear numerical solution (solid black
curves) for various interfacial quantities. \label{fig:k_lambda-mode-1}}
\end{figure}

\begin{figure}
\noindent \begin{centering}
\textbf{(a)} \enskip{}\includegraphics[width=0.6\textwidth]{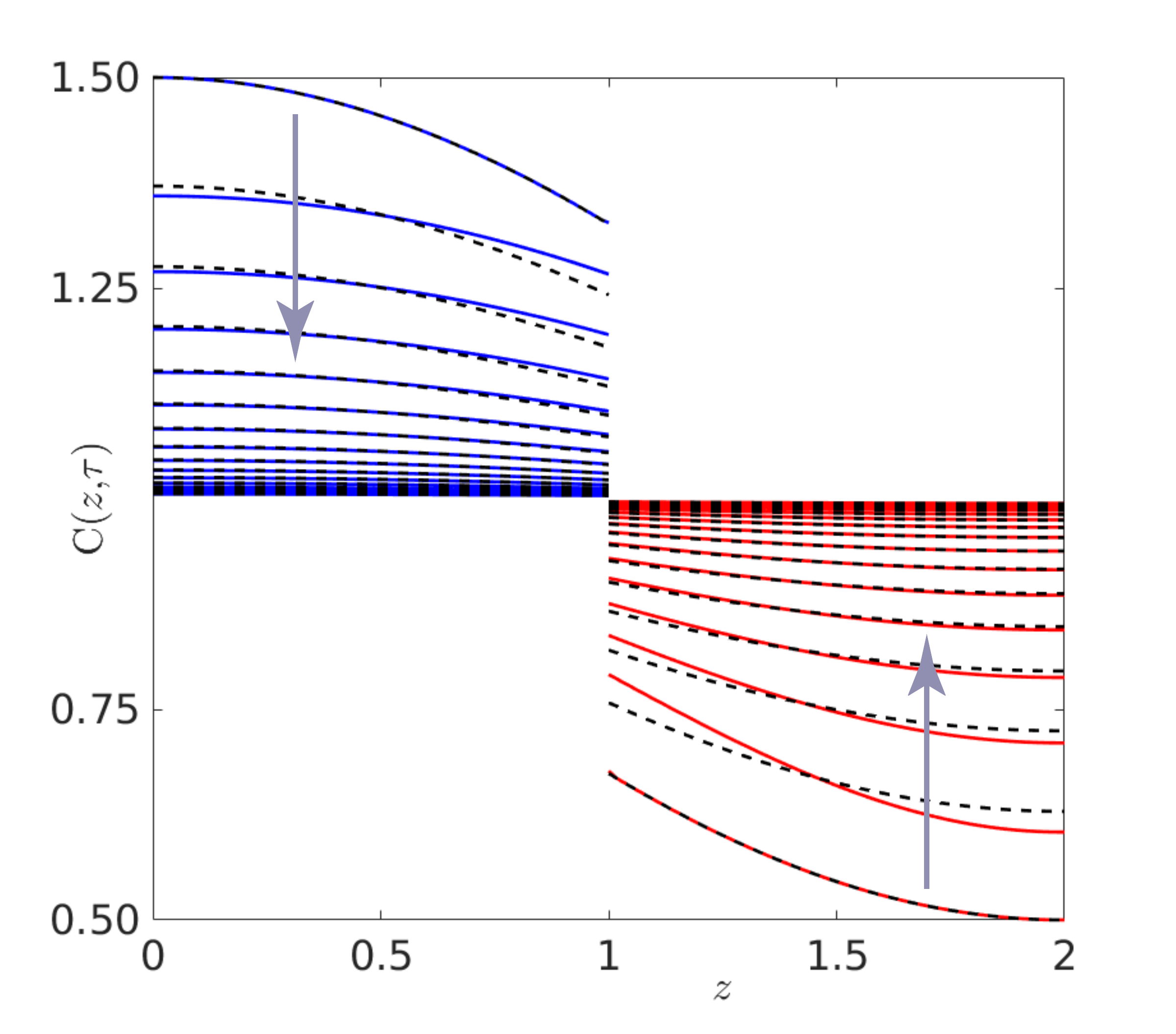}
\par\end{centering}
\bigskip{}

\noindent \begin{centering}
\textbf{(b)} \enskip{}\includegraphics[width=0.6\textwidth]{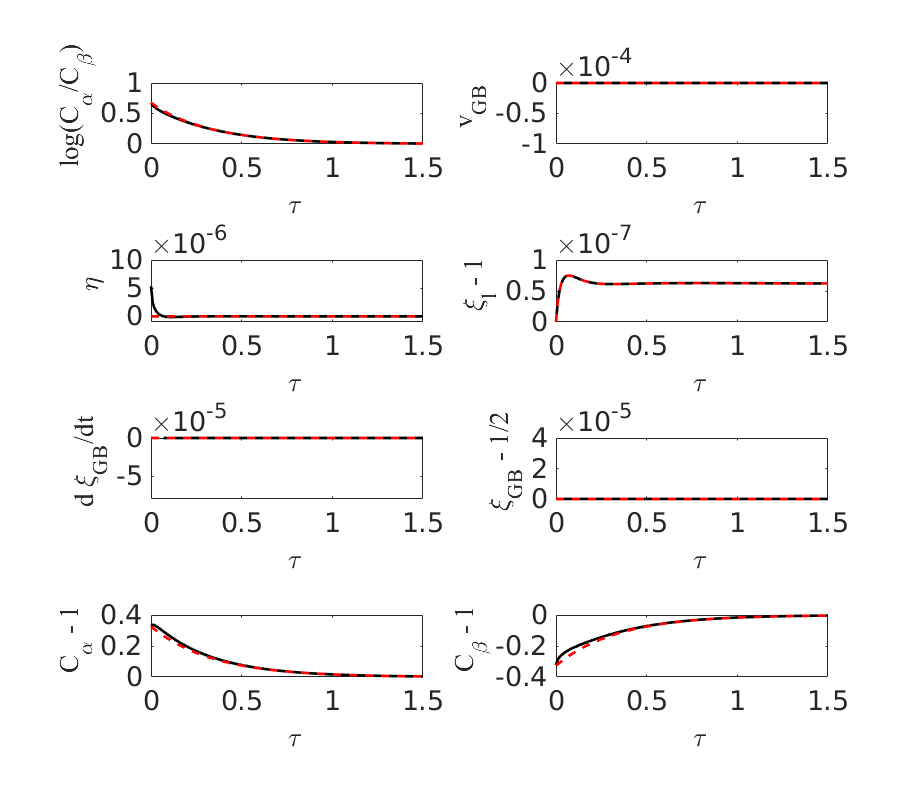}
\par\end{centering}
\caption{(a) Comparison of the $\kappa_{\lambda}$ normal mode solution (dashed
black curves) and the nonlinear numerical solution (solid blue and
red curves) for the vacancy concentration fields in each grain. The
concentration profiles are shown at times $\tau$ ranging from 0 to
1.5 at 0.1 increments.
The arrows indicate the direction of time. The model parameters are
$\lambda=0$, $\lambda_{t}=2\times10^{-4}$ and $\rho=10^{-4}$ (thus
$Z_{\rho}=1/4$ and $Z_{\lambda}=1$). (b) Comparison of the same
normal mode solution (dashed red curves) and nonlinear numerical solution
(solid black curves) for various interfacial quantities.\label{fig:k_lambda-mode-2}}
\end{figure}

The $\kappa_{\lambda}$ mode was implemented by creating the initial
vacancy distribution according to Eqs.(\ref{eq:Profile_lambda_1})
and (\ref{eq:profile_lambda_2}) with $\tau=0$ and $\hat{C}=0.5$
(Figs.~\ref{fig:k_lambda-mode-1} and \ref{fig:k_lambda-mode-2}).
We again choose $\rho=10^{-4}$ and thus $Z_{\rho}=1/4$. The parameter
$Z_{\lambda}$ depends on the combination $2\lambda+\lambda_{t}$.
Instead of choosing one particular value of this combination, two
different cases were tested, namely, $\lambda=10^{-4}$ and $\lambda_{t}=0$,
and $\lambda=0$ and $2\times10^{-4}$. Although both cases have the
same $Z_{\lambda}=1$, the situations are physically different. In
the first case, the GB is mobile while diffusion across the GB is
prohibited. The excess vacancies diffuse from grain $\alpha$ toward
the left-hand side of the boundary, where they are annihilated. By
the symmetry of the equations, the right-hand side of the boundary
simultaneously injects the same amount of vacancies into grain $\beta$.
As a result, grain $\alpha$ shrinks while grain $\beta$ expands,
keeping the same size of the system and zero lattice velocity in grain
$\beta$ (Fig.~\ref{fig:k_lambda-mode-1}). The GB migrates toward
grain $\alpha$ while the vacancy concentration jump at the boundary
decreases and eventually closes. In the second case, the vacancies
diffuse across the GB, from grain $\alpha$ to grain $\beta$, which
again results in reducing and eventually closing the concentration
gap (Fig.~\ref{fig:k_lambda-mode-2}). However, while there is a
driving force for GB migration, the boundary does not move due to
its zero mobility coefficient ($\lambda=0$). In both cases, the time
evolution of the vacancy concentration is similar and relatively close
to the respective normal mode solution (\ref{eq:Profile_lambda_1})-(\ref{eq:profile_lambda_2}).
Similar tests with a small perturbation ($\hat{C}=0.5\times10^{-4}$)
show much closer agreement between the nonlinear model and the respective
normal mode solution. (see Figs.~3 and 4 in the Supplementary Material
\citep{Supplementary-Material}). These calculations demonstrate that,
in spite of the drastic difference in the $\lambda$ and $\lambda_{t}$
parameters and the operation of different physical mechanisms, the
vacancy relaxation kinetics is in both cases similar and is primarily
governed by the parameters $Z_{\rho}$ and $Z_{\lambda}$.

\subsubsection{Mixed-mode cases}

Further calculations were performed in the mixed-mode regime by creating
an initial vacancy distribution different from that in either normal
mode. Namely, the initial vacancy concentration profile was chosen
to be
\begin{equation}
C_{\alpha}=1,\qquad0\leq\xi\leq1/2,\label{eq:map1-1}
\end{equation}
\begin{equation}
C_{\beta}=\dfrac{3}{2}+\dfrac{3}{4}\left(4\xi-3\right)-\dfrac{1}{4}\left(4\xi-3\right)^{3},\qquad1/2\leq\xi\leq1.\label{eq:map1-1-1}
\end{equation}
By these equations, the vacancy concentration is at
equilibrium in grain $\alpha$ ($C_{\alpha}\equiv1$), twice the
equilibrium value at the right end of the sample ($C_{\beta}=2$ at
$\xi=1$), and tends
smoothly to the equilibrium value when approaching the GB from grain
$\beta$. Since this function is neither even nor odd with respect
to the GB position $\xi=1/2$, the vacancy relaxation process does
not have to follow a particular normal mode. The vacancy concentration
jump and concentration gradient at the boundary are initially zero.

Table \ref{tab:Parameters} summarizes eight combinations of the
dimensionless model parameters that were tested, where each parameter
was taken to be either very large or very small.
The orders of magnitude of these dimensionless parameters were chosen
based on estimates of the possible ranges of the physical parameters
on which they depend. The accompanying online Supplementary file
\citep{Supplementary-Material} contains a collection of the vacancy
concentration profiles and various interface properties as functions
of time for all eight cases. Four of
the most representative cases will be discussed below.

\begin{figure}
\noindent \begin{centering}
\textbf{(a)} \enskip{}\includegraphics[width=0.6\textwidth]{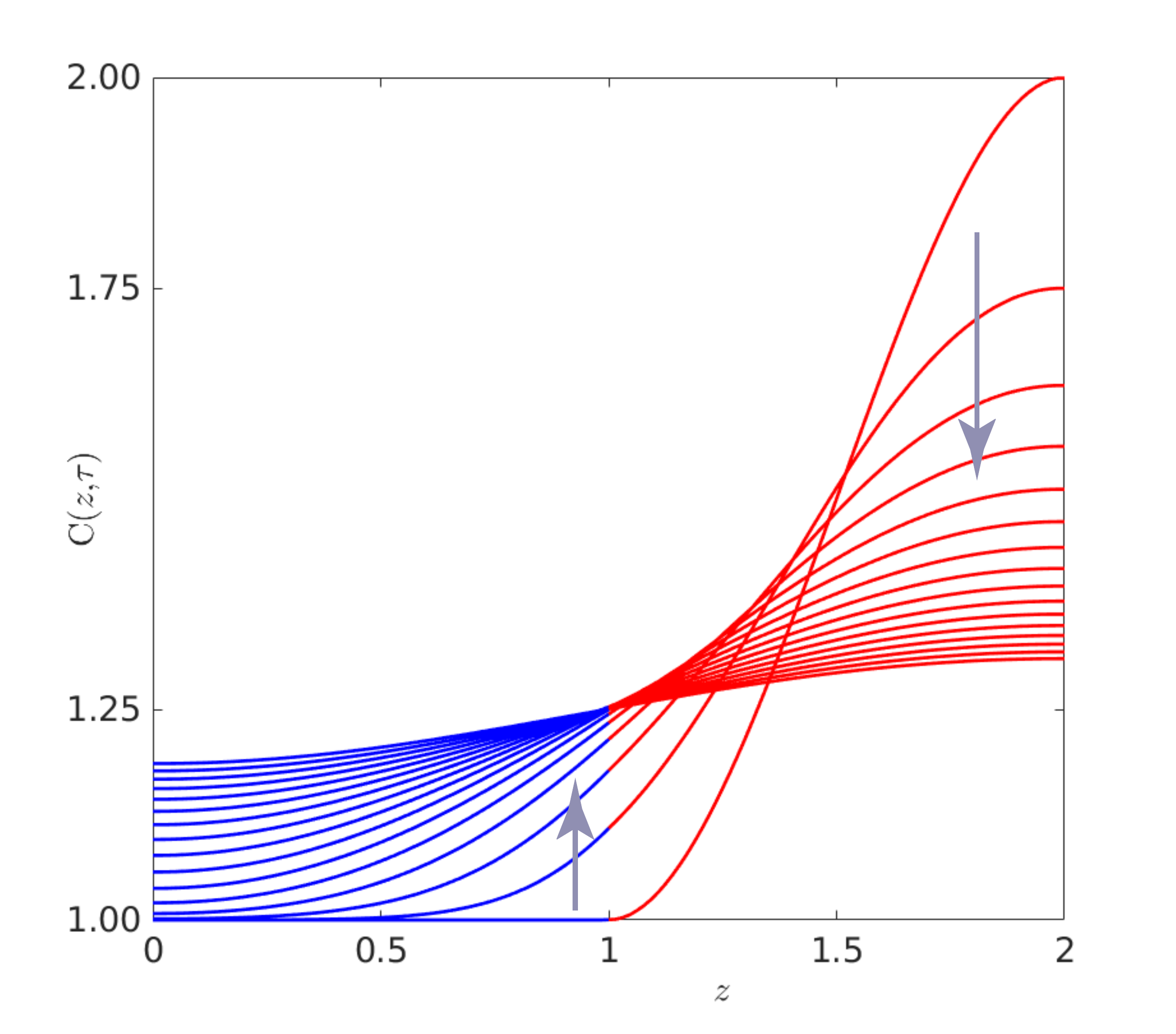}
\par\end{centering}
\bigskip{}

\noindent \begin{centering}
\textbf{(b)} \enskip{}\includegraphics[width=0.6\textwidth]{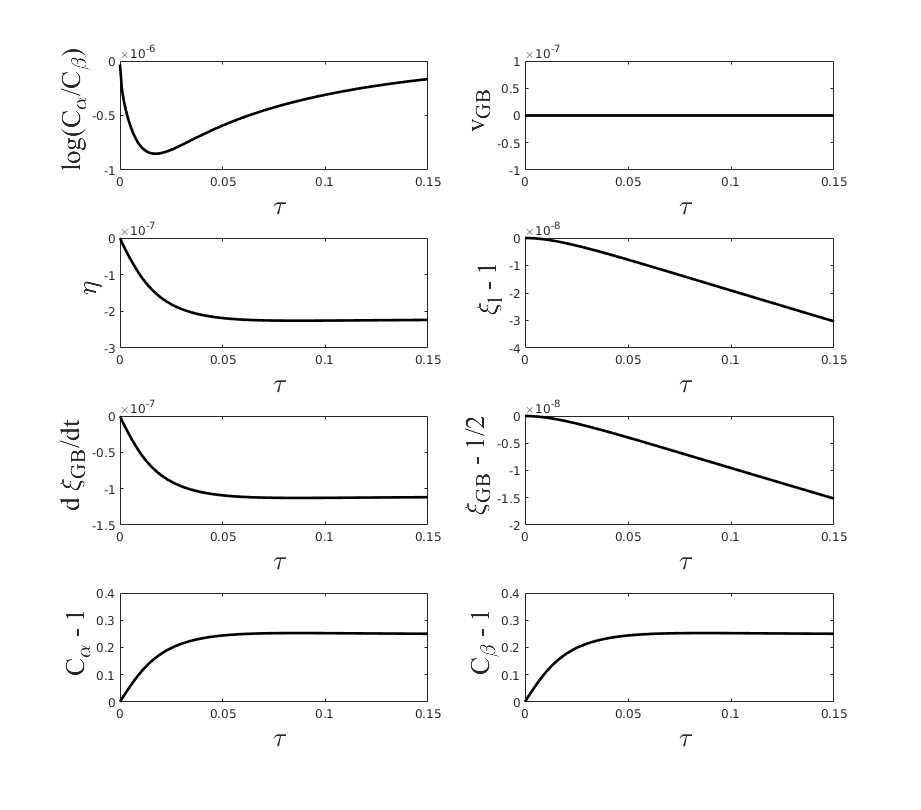}
\par\end{centering}
\caption{Vacancy concentration profiles (a) and various interfacial quantities.
(b) obtained by numerical solution of the model. The concentration
profiles are shown at times $\tau$ ranging from 0 to 1.5 at 0.1 increments.
The arrows indicate the direction of time. The model parameters correspond
to case 4 in Table \ref{tab:Parameters}.\label{fig:case_4}}
\end{figure}

In the first example, corresponding to case 4 in Table \ref{tab:Parameters},
$\lambda$ and $\rho$ are chosen to be small, meaning that the GB
is highly immobile and virtually incapable of vacancy absorption.
At the same time, $\lambda_{t}$ is large, so that the GB does not
pose any significant resistance to vacancy diffusion. In effect, this
situation is almost equivalent to the absence of a GB and thus of
any vacancy sinks/sources in the system. The relaxation process reduces
to diffusion-controlled vacancy redistribution between the two grains.
Since the vacancies are almost conserved, the process is expected
to result in uniform vacancy distribution across the system with the
concentration close to $C_{\alpha}=C_{\beta}=1.25$ (obtained by the
averaging of Eqs.(\ref{eq:map1-1}) and (\ref{eq:map1-1-1})). The
numerical results presented in Fig.~\ref{fig:case_4} generally agree
with this scenario while also showing some very small lattice and
GB velocities and a small concentration gap at the GB arising due
to the nonzero values of $\lambda$ and $\rho$ and a finite value
of $\lambda_{t}$.

\begin{figure}
\noindent \begin{centering}
\textbf{(a)} \enskip{}\includegraphics[width=0.6\textwidth]{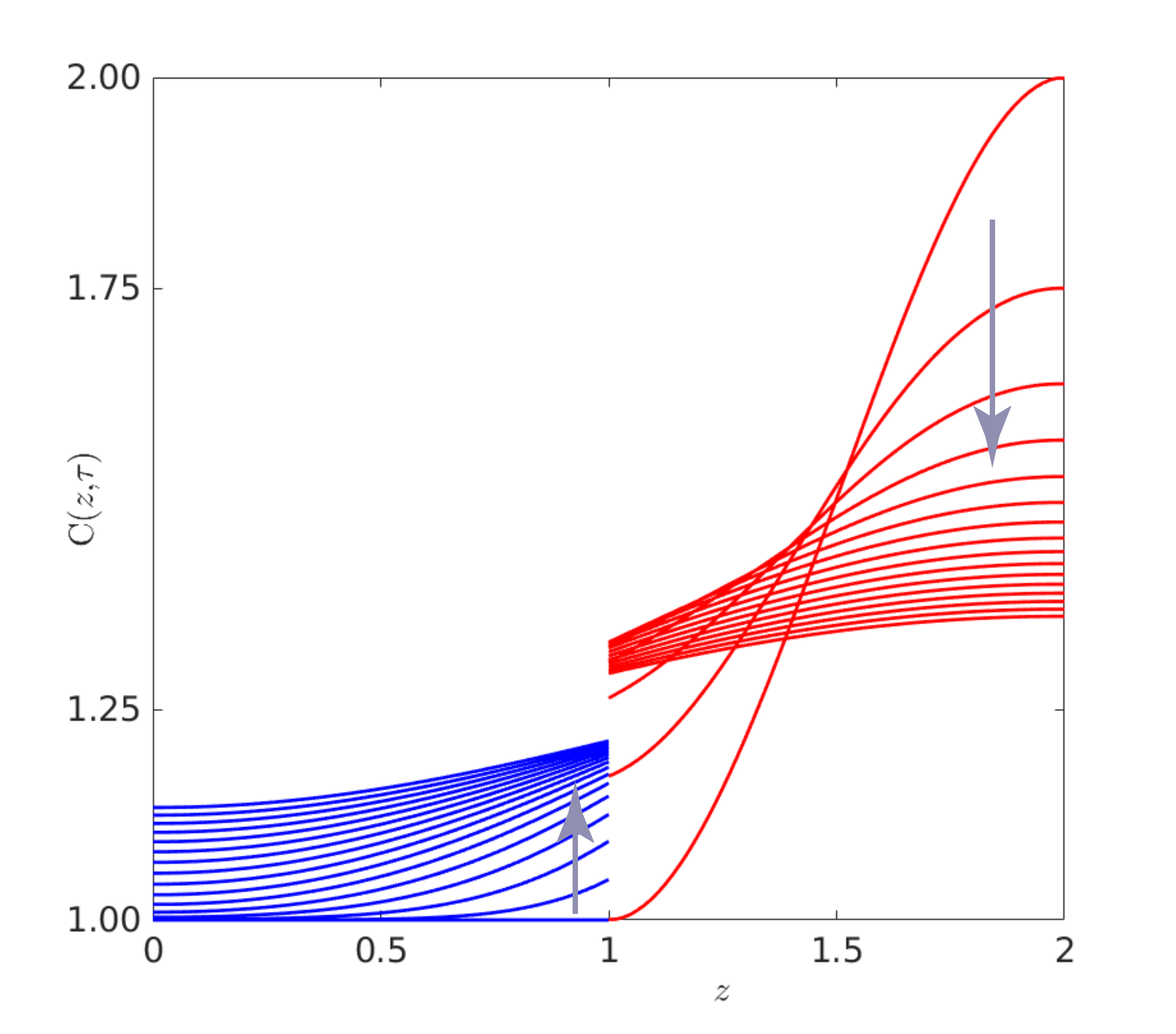}
\par\end{centering}
\bigskip{}

\noindent \begin{centering}
\textbf{(b)} \enskip{}\includegraphics[width=0.6\textwidth]{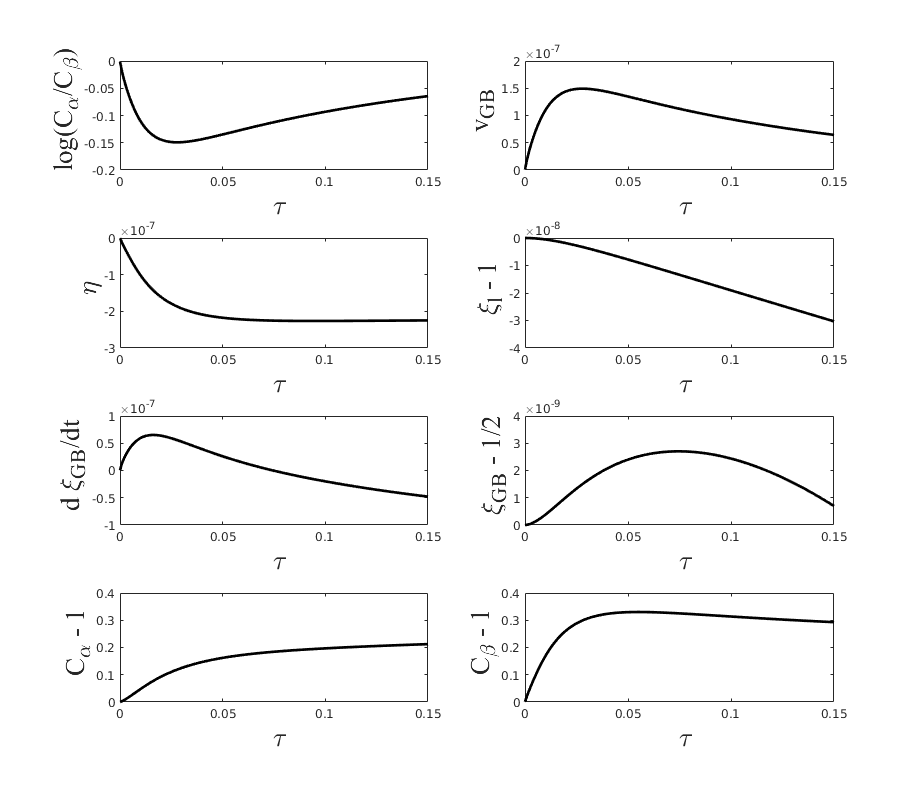}
\par\end{centering}
\caption{Vacancy concentration profiles (a) and various interfacial quantities.
(b) obtained by numerical solution of the model. The concentration
profiles are shown at times $\tau$ ranging from 0 to 1.5 at 0.1 increments.
The arrow indicates the direction of time. The model parameters correspond
to case 8 in Table \ref{tab:Parameters}.\label{fig:case_8}}
\end{figure}

In the second example, all three parameters $\lambda$, $\lambda_{t}$
and $\rho$ are small (case 8 in Table \ref{tab:Parameters}). The
difference from the previous example is that the GB is now a strong
diffusive barrier to the vacancies. To the first approximation, the
grains can be treated as isolated from each other. Then the vacancy
concentration in grain $\alpha$ should remain $C_{\alpha}\equiv1$
while in grain $\beta$ it should tend to even out and eventually
becomes uniform at $C_{\beta}=1.5$. This process should create a
vacancy concentration gap $\Delta C=C_{\beta}-C_{\alpha}=0.5$ at
the boundary. The numerical solution shows that a large gap does initially
arise. However, the vacancies soon begin to leak into grain $\alpha$,
reducing the gap and, in the long run, closing it (Fig.~\ref{fig:case_8}).
The vacancy concentration tends to level at the conservation-dictated
value of 1.25. A notable feature of this process is that the gap creates
a driving force for GB migration, as evident from the equation
\begin{equation}
\eta_{GB}=-\lambda\ln\dfrac{C_{\alpha}}{C_{\beta}}\label{eq:168-1}
\end{equation}
obtained by combining Eqs.(\ref{eq:43a}) and (\ref{eq:168}). Although
the mobility $\lambda$ is small, the GB still moves with a velocity
$\eta_{GB}$ that is orders of magnitude higher than in the previous
example where the gap was small. This example demonstrates the physical
phenomenon of vacancy-driven GB migration.

\begin{figure}
\noindent \begin{centering}
\textbf{(a)} \enskip{}\includegraphics[width=0.6\textwidth]{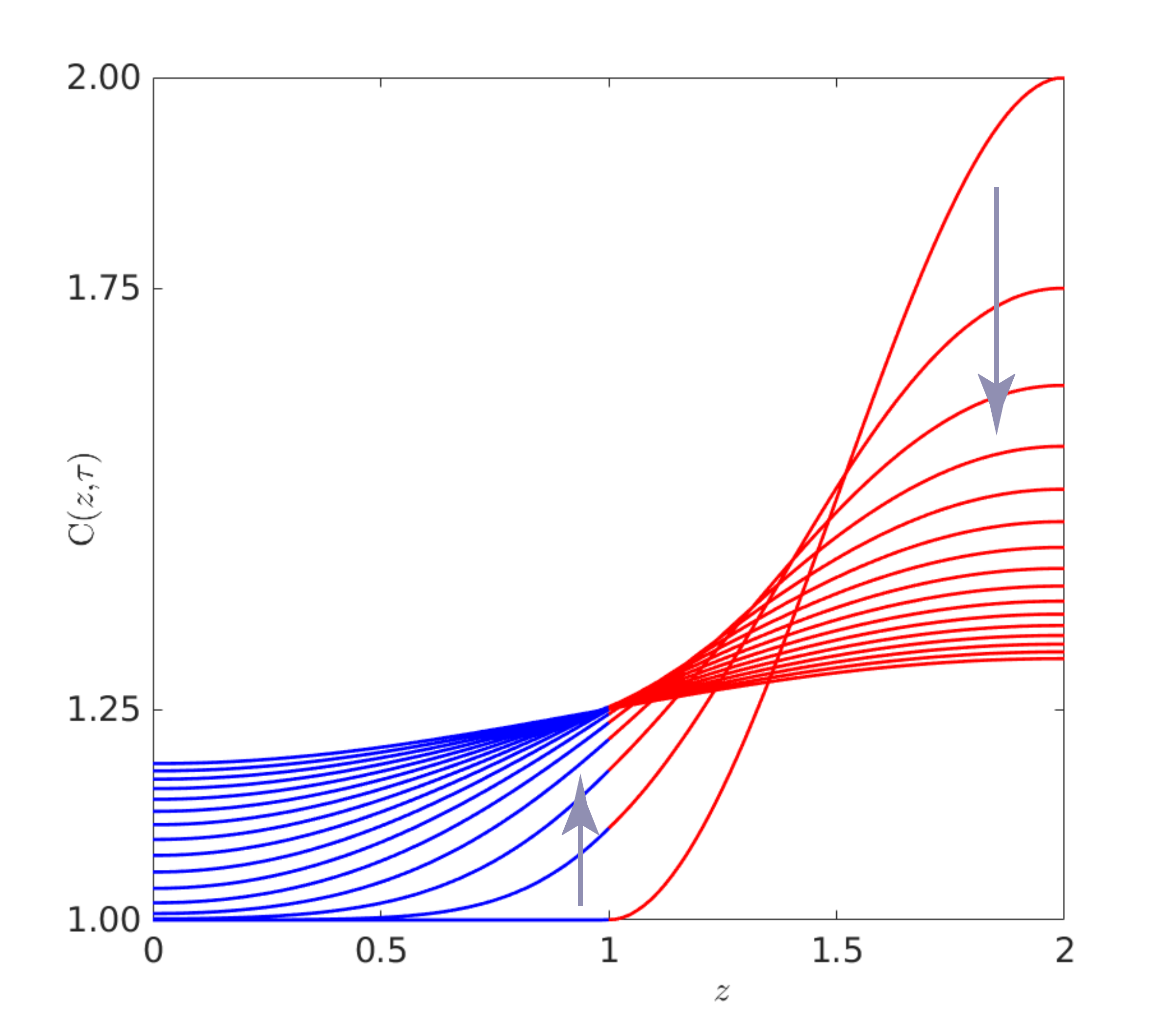}
\par\end{centering}
\bigskip{}

\noindent \begin{centering}
\textbf{(b)} \enskip{}\includegraphics[width=0.6\textwidth]{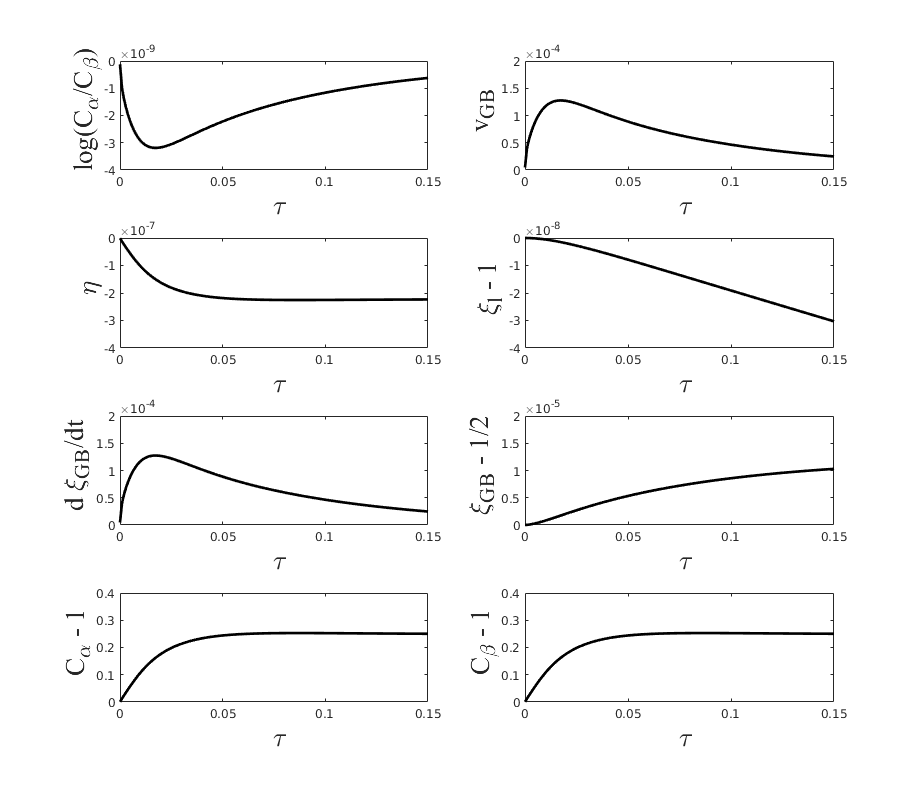}
\par\end{centering}
\caption{Vacancy concentration profiles (a) and various interfacial quantities.
(b) obtained by numerical solution of the model. The concentration
profiles are shown at times $\tau$ ranging from 0 to 1.5 at 0.1 increments.
The arrows indicate the direction of time. The model parameters correspond
to case 7 in Table \ref{tab:Parameters}.\label{fig:case_7}}
\end{figure}

In the third example, corresponding to case 7 in Table \ref{tab:Parameters},
the vacancy-driven GB migration occurs much faster than in the previous
example. We keep the same small values of $\lambda_{t}$ and $\rho$
(slow diffusion across the GB and negligible vacancy absorption),
but the GB mobility $\lambda$ is now high. Interestingly, the concentration
gap at the GB becomes much smaller (Fig.~\ref{fig:case_7}). Nevertheless,
due to the high GB mobility this small gap is still capable of driving
the boundary motion into grain $\beta$ with a much higher velocity
than in the previous examples.

It should be noted that the relation between the vacancy concentration
gap and the GB motion is rather convoluted. While the GB motion is
driven by the gap, it also strongly reduces the gap. To understand
the underlying mechanism, suppose a large gap initially exists due
to the diffusion barrier at the GB. Driven by this gap, the boundary
moves into the grain in which the vacancy concentration is smaller
(cf.~Eq.(\ref{eq:168-1})). The gap is then left behind, inside the
advancing grain, and can now smooth out by lattice diffusion. In the
new position, the GB starts creating a new concentration gap by acting
as a diffusion barrier, until the gap becomes large enough to move
the boundary into a new position, leaving the gap behind for smoothing
by lattice diffusion. In reality the process is continuous rather
than incremental, but the mechanism is the same. The process eventually
reaches a kinetic balance between the gap-driven GB motion and the
motion-induced gap suppression. The remarkable result of this gap-motion
coupling is that, in spite of the existence of the strong diffusion
barrier at the GB, the vacancy composition profiles look very similar
to those in the first example (cf.~Fig.~\ref{fig:case_4}), in which
the boundary did not impose any diffusion barrier but was immobile.
This similarity confirms the prominent role of the parameter combination
$2\lambda+\lambda_{t}$ in this model, as suggested by the normal
mode analysis. In the two cases compared here, this combination is
large but for very different reasons: $2\lambda\ll\lambda_{t}$ in
one case and $2\lambda\gg\lambda_{t}$ in the other. Note also that
the profiles gradually develop a shape that is nearly odd with respect
to the GB position, which is a feature the $\kappa_{\lambda}$ normal
mode. Thus, the gap-motion coupling provides the mechanistic explanation
of why $\lambda$ and $\lambda_{t}$ control the vacancy relaxation
process predominantly in this combination and not separately.

\begin{figure}
\noindent \begin{centering}
\textbf{(a)} \enskip{}\includegraphics[width=0.6\textwidth]{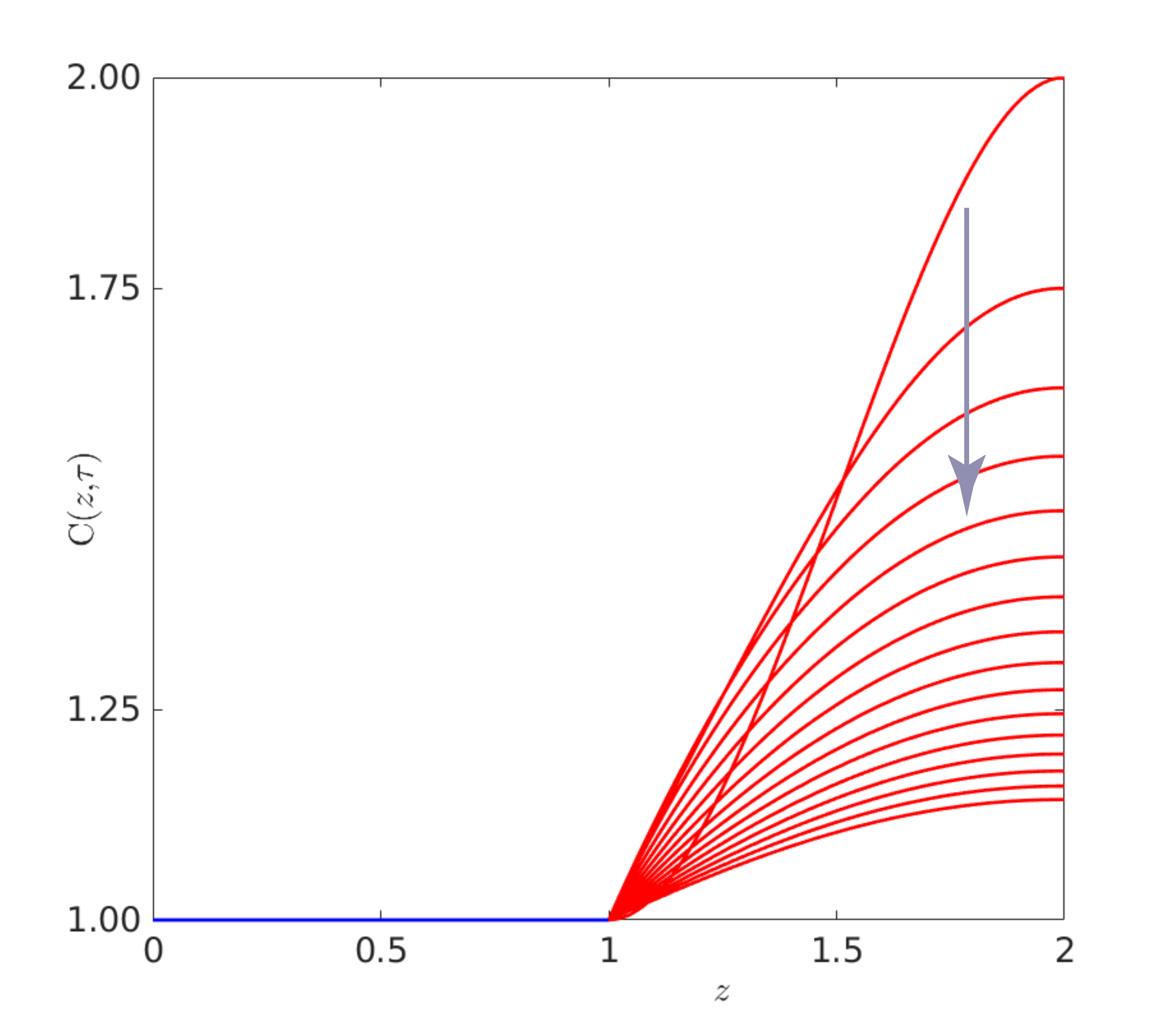}
\par\end{centering}
\bigskip{}

\noindent \begin{centering}
\textbf{(b)} \enskip{}\includegraphics[width=0.6\textwidth]{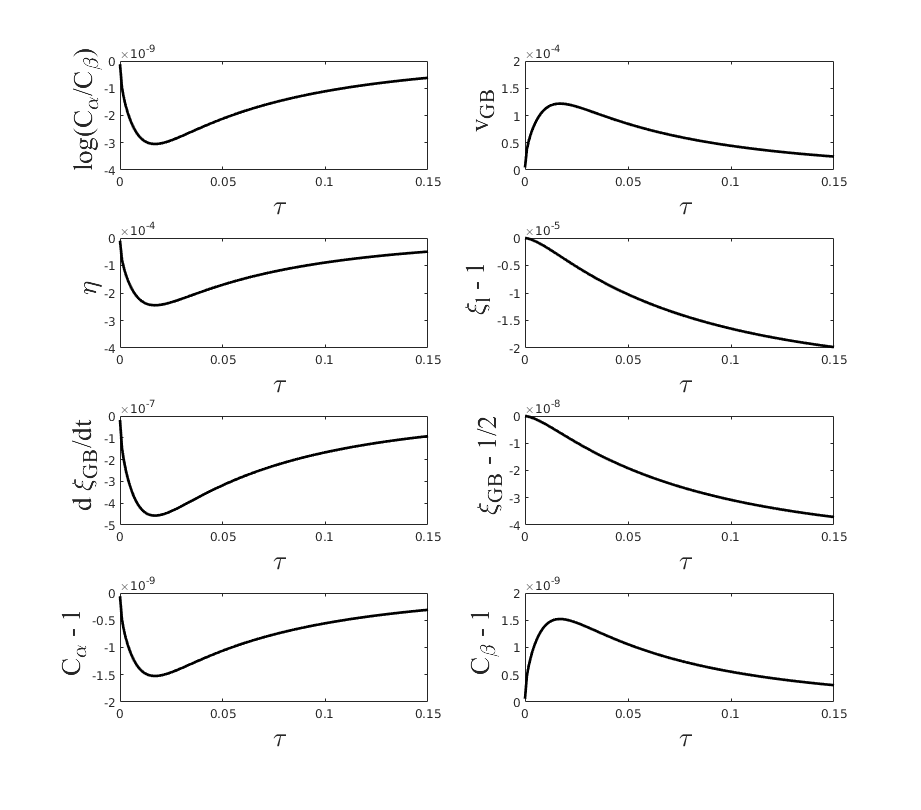}
\par\end{centering}
\caption{Vacancy concentration profiles (a) and various interfacial quantities.
(b) obtained by numerical solution of the model. The concentration
profiles are shown at times $\tau$ ranging from 0 to 1.5 at 0.1 increments.
The arrow indicates the direction of time. The model parameters correspond
to case 1 in Table \ref{tab:Parameters}.\label{fig:case_1}}
\end{figure}

Finally, example four features the effect of vacancy absorption at
the boundary. The model parameters correspond to case 1 in Table \ref{tab:Parameters},
in which the vacancy diffusion across the boundary is fast (large
$\lambda_{t}$), the GB is highly mobile (large $\lambda$), and it
acts as a powerful sink/source of vacancies (large $\rho$). The GB
maintains a nearly perfect equilibrium concentration at all times
(Fig.~\ref{fig:case_1}). This is a standard assumption in most models
describing GBs as vacancy sinks. The vacancies diffusing toward the
boundary from grain $\beta$ are eliminated at the boundary and do
not penetrate into grain $\alpha$. As a result, grain $\beta$ shrinks
while grain $\alpha$ remains intact in both its size and the vacancy
concentration. Since the lattice velocity $\eta$ is negative and
much larger in magnitude than $d\xi_{GB}/d\tau$, the process can
be described as GB migration into grain $\beta$ with the velocity
$\eta_{GB}=-\eta/2$ driven by the small but non-negligible vacancy
concentration jump.

\section{Conclusions\label{sec:Conclusions}}

We have applied a sharp-interface model to study the interactions
of a planar GB with non-equilibrium vacancies in the presence of mechanical
stresses. By contrast to most of the existing models, the present
model includes the vacancy generation and absorption processes (or
equivalently, lattice site generation/absorption) by the GB. The model
predicts a set of coupled kinetic processes that can occur at the
GB, including vacancy diffusion toward or away from the boundary,
vacancy diffusion across the GB, the GB migration process, and the
vacancy generation/absorption at the GB. The rates of these processes
are linked to the respective thermodynamic forces, which were previously
\citep{Mishin:2015ad} identified from the equation for the total
free energy rate of decrease.

Analysis of the model and the numerical calculations reported here
reveal a number of effects arising due to the coupling among the different
processes. In particular, the GB motion accelerates the absorption
of over-saturated vacancies and/or generation of new vacancies into
the grains if their concentration is below equilibrium. If the vacancy
concentration is different on either side of the GB, a driving force
arises causing GB motion into the grain with the larger concentration.
At the same time, the GB motion reduces the concentration jump across
the GB and thus the driving force for its migration. The vacancy concentration
jump is also influenced by the vacancy generation/absorption process.
The different vacancy concentrations result in different driving forces
for the vacancy generation or absorption, and thus different generation/absorption
rates. The difference in the generation/absorption rates on either
side of the GB causes its migration into the grain with faster absorption
or slower generation. All these processes are influenced by the rate
of vacancy diffusion across the GB. Slow cross-boundary diffusion
creates a vacancy concentration jump, which creates a driving force
for the GB motion, which in turn affects the rate of vacancy generation
and absorption.

Given this coupling among the kinetic processes involved, any meaningful
model of vacancy-GB interactions must include all of these processes,
describing them by a set of coupled equations. The model presented
here achieves this goal and identifies the key combinations of the
material parameters that can be utilized to control the vacancy-GB
interaction process for various applications. Potential applications
of the model include radiation damage healing, diffusional creep,
irradiation
creep,
and solid-state sintering \citep{Herring:1951aa,Herring53,Kazaryan:1999aa,Wang:2006aa,Li-2011,Abdeljawad:2019aa}.

The present version of the model is based on many simplifying assumptions
and approximations, some of which can be lifted in the future. For
example, the model can be generalized to multi-component solid solutions
containing both vacancies and interstitials. This should allow one
to describe the formation of segregation atmospheres and denuded zones
near GBs. Although the GB studied here was planar, extension to curved
GBs is also possible along the lines discussed in \citep{Mishin:2015ad}.
it should also be noted that some of the kinetic parameters of the
model can be obtained by atomistic calculations, including the lattice
and grain boundary diffusion coefficients as well as the grain boundary
mobility. Obtaining this information from atomistic studies of crystallographically
specific grain boundaries, and using the present model, could provide
a better understanding of the effect of bicrystallography and atomic
structure of grain boundaries on their capacity to absorb or generate
point defects.

\vspace{0.15in}

\textbf{Acknowledgement:} Y.M was supported by the National Institute
of Standards and Technology, Material Measurement Laboratory, the
Materials Science and Engineering Division.

%\bibliographystyle{/Users/ymishin/YURI/Bibliography/ActaMatnew}
%\bibliography{/Users/ymishin/YURI/Bibliography/literat}

\begingroup
\raggedright

\endgroup

\newpage\clearpage\setcounter{table}{0}
\renewcommand{\thetable}{\Alph\Part A\arabic{table}}

\part*{Appendix A: Table of notation}

\renewcommand{\thetable}{\arabic{table}}{}%
\begin{longtable}[c]{ll>{\raggedright}p{0.8\textwidth}}
\caption{List of notation
in the order of appearance in the paper. Indices $\alpha$ and
$\beta$ label the two grains separated by a grain boundary.}
\tabularnewline
\hline 
Notation &  & \qquad{}\qquad{}\qquad{}\qquad{}Meaning\tabularnewline
\hline 
$\mathbf{v}_{L}^{\alpha}$
and $\mathbf{v}_{L}^{\beta}$ &  & Lattice velocities
in the grains with respect to laboratory reference frame.\tabularnewline
$\mathbf{J}_{L}^{\alpha}$
and $\mathbf{J}_{L}^{\beta}$ &  & Atomic diffusion
fluxes in the grains with respect to the lattice reference frames.\tabularnewline
$\mathbf{J}_{b}$  &  & 2D atomic diffusion
flux in grain boundary.\tabularnewline
$\mathbf{J}_{n}$ &  & Atomic diffusion
flux across grain boundary.\tabularnewline
$\mathbf{n}^{\alpha}$ &  & Normal to grain boundary
pointing from grain $\alpha$ into grain $\beta$.\tabularnewline
$\mathbf{\boldsymbol{\sigma}_{\alpha}}$
and $\mathbf{\boldsymbol{\sigma}_{\beta}}$ &  & Stress tensors in
the grains.\tabularnewline
$\omega_{\alpha}$
and $\omega_{\beta}$ &  & Grand-canonical potentials
per unit volume in the grains.\tabularnewline
$f_{s}^{\alpha}$
and $f_{s}^{\beta}$  &  & Free energies per
lattice site in the grains.\tabularnewline
$\Omega_{\alpha}$
and $\Omega_{\beta}$  &  & Atomic volumes per
lattice site in the grains.\tabularnewline
$M_{\alpha}$ and
$M_{\beta}$ &  & Diffusion potentials
of atoms relative to the vacancies in the grains.\tabularnewline
$c_{\alpha}$ and
$c_{\beta}$ &  & Atomic concentrations
(fractions of occupied lattice sites) in the grains.\tabularnewline
$L_{\alpha}$ and
$L_{\beta}$ &  & Diffusion kinetic
coefficients in the grains.\tabularnewline
$\mathbf{x},t$ &  & Position and time.\tabularnewline
$\mathbf{A}_{\alpha}$
and $\mathbf{A}_{\beta}$  &  & General vector fields
in the grains extrapolated to the grain boundary.\tabularnewline
$\left[\mathbf{A}\right]\equiv\mathbf{A}_{\alpha}-\mathbf{A}_{\beta}$ &  & Jump
of the vector fields across the grain boundary.\tabularnewline
$\left\langle \mathbf{A}\right\rangle \equiv(\mathbf{A}_{\alpha}+\mathbf{A}_{\beta})/2$ &  & Average 
value of the vector fields across the grain boundary.\tabularnewline
$L_{t}$ and $L_{p}$ &  & Kinetic coefficients
for trans-boundary and intra-boundary diffusion.\tabularnewline
$\nu_{\alpha}$ and
$\nu_{\beta}$ &  & Numbers of atoms
added to the grains per unit area and unit time.\tabularnewline
$\nabla_{b}$ &  & 2D gradient in grain
boundary plane.\tabularnewline
$\mathbf{v}_{b}$  &  & Grain boundary velocity
relative to laboratory reference frame.\tabularnewline
$\mathbf{s}$ &  & Grain boundary traction
vector.\tabularnewline
$\bot$ and $\Vert$ &  & Subscripts for normal
and tangential components.\tabularnewline
$R$ &  & Kinetic coefficient
for vacancy creation and annihilation in the grain boundary.\tabularnewline
$L_{GB}$ &  & Kinetic coefficient
for grain boundary mobility.\tabularnewline
$\beta$ &  & Shear coupling factor.\tabularnewline
$v_{GB}$ &  & Grain boundary velocity
with respect to the mean lattice velocity of the grains.\tabularnewline
\textbf{$\mathbf{t}$} &  & Direction of shear-coupled
grain boundary motion.\tabularnewline
$K_{s}$ &  & Kinetic coefficient
for grain boundary sliding.\tabularnewline
$\Delta\Omega_{v}$ &  & Vacancy relaxation
volume.\tabularnewline
$\boldsymbol{\varepsilon}^{0}$ &  & Stress-free strain
due to vacancies.\tabularnewline
$\Omega^{\prime}$  &  & Atomic volume in
the reference state of strain.\tabularnewline
$\mathbf{S}$ &  & Tensor of elastic
compliances.\tabularnewline
$\sigma_{h}$ &  & Hydrostatic part
of the stress tensor.\tabularnewline
$M^{*}$ &  & Equilibrium diffusion
potential.\tabularnewline
$c^{*}$ &  & Equilibrium atomic
concentration in stress-free grains.\tabularnewline
$kT$ &  & Thermal energy ($T$
temperature, $k$ Boltzmann's constant).\tabularnewline
$c_{v}$ &  & Vacancy concentration
(fraction of vacant lattice sites).\tabularnewline
$c_{v}^{*}$ &  & Equilibrium vacancy
concentration.\tabularnewline
$D$ &  & Atomic diffusion
coefficient.\tabularnewline
$x_{GB}$ &  & Grain boundary position.\tabularnewline
$l(t)$ &  & Bicrystal length
as a function of time.\tabularnewline
$l_{0}$ &  & Initial bicrystal
length.\tabularnewline
$E$, $\nu$ &  & Young modulus and
Poisson's ratio of the material.\tabularnewline
$L_{t}^{\prime}$ &  & Scaled kinetic coefficient
of trans-boundary diffusion.\tabularnewline
$\tau$ &  & Dimensionless time.\tabularnewline
$\xi$ &  & Dimensionless coordinate.\tabularnewline
$\xi_{l}$ &  & Dimensionless bicrystal
length.\tabularnewline
$\xi_{GB}$ &  & Dimensionless grain
boundary position.\tabularnewline
$\eta$ &  & Dimensionless lattice
velocity.\tabularnewline
$s$ &  & Dimensionless stress.\tabularnewline
$\lambda_{t}$ &  & Dimensionless kinetic
coefficient of trans-boundary diffusion.\tabularnewline
$\lambda$ &  & Dimensionless grain
boundary mobility.\tabularnewline
$\rho$ &  & Dimensionless kinetic
coefficient for vacancy creation and annihilation in the grain boundary.\tabularnewline
$b$ &  & Dimensionless vacancy
relaxation volume.\tabularnewline
$\hat{C}$  &  & Dimensionless amplitude
of the excess of the vacancy concentration.\tabularnewline
$\kappa$, $\kappa_{\rho}$,
$\kappa_{\lambda}$ &  & Relaxation wave numbers
in the linear stability analysis.\tabularnewline
$Z_{\rho}$, $Z_{\lambda}$ &  & Dimensionless parameters
governing the relaxation modes in the linear stability analysis.\tabularnewline
\hline 
\end{longtable}\lyxadded{Yuri Mishin}{Mon Oct 21 15:22:53 2019}{\newpage\clearpage}

\part*{Appendix B: Linear stability analysis}

The governing equations (\ref{eq:165})-(\ref{eq:168}) with $s=0$
can be linearized around the equilibrium state with $C_{\alpha}=C_{\beta}=1$,
$\xi_{GB}=1/2$, $\eta=0$ and $\xi_{l}=1$. The linearized equations
for the perturbed vacancy concentrations $c_{a}(\xi,\tau)$ and $c_{b}(\xi,\tau)$
are
\begin{equation}
\dfrac{\partial c_{a}}{\partial\tau}=\dfrac{\partial^{2}c_{a}}{\partial\xi^{2}},\:\:\:\mathrm{0\leq\xi\leq1/2},\label{eq:175}
\end{equation}
\begin{equation}
\dfrac{\partial c_{b}}{\partial\tau}=\dfrac{\partial^{2}c_{b}}{\partial\xi^{2}},\:\:\:\mathrm{1/2\leq\xi\leq1,}\label{eq:176}
\end{equation}
with the boundary conditions
\[
\dfrac{\partial c_{a}}{\partial\xi}(0,\tau)=0,\:\:\:\dfrac{\partial c_{b}}{\partial\xi}(1,\tau)=0,
\]
The linearized boundary conditions at the GB ($\xi=1/2$) are
\begin{equation}
-2\dfrac{d\xi_{GB}}{d\tau}+\eta+c_{v}^{*}\left(\dfrac{\partial c_{a}}{\partial\xi}+\dfrac{\partial c_{b}}{\partial\xi}\right)=-\lambda_{t}\left(c_{a}-c_{b}\right),\label{eq:177}
\end{equation}
\begin{equation}
\eta+c_{v}^{*}\left(\dfrac{\partial c_{a}}{\partial\xi}-\dfrac{\partial c_{b}}{\partial\xi}\right)=0,\label{eq:178}
\end{equation}
\begin{equation}
\eta=-\rho\dfrac{c_{a}+c_{b}}{2},\label{eq:179}
\end{equation}
\begin{equation}
\dfrac{d\xi_{GB}}{d\tau}=\dfrac{1}{2}\eta-\lambda\left(c_{a}-c_{b}\right).\label{eq:180}
\end{equation}

We seek a normal mode solution of the linearized equations in the
form
\[
c_{a}(\xi,\tau)=C_{a}(\xi)\exp(-\kappa^{2}\tau),\:\:\:c_{b}(\xi,\tau)=C_{b}(\xi)\exp(-\kappa^{2}\tau),
\]
\[
\xi_{GB}(\tau)=\dfrac{1}{2}+X_{GB}\exp(-\kappa^{2}\tau),\:\:\:\eta(\tau)=V\exp(-\kappa^{2}\tau),
\]
where we anticipate damped solutions with a real, positive decay rate
$\kappa^{2}>0$. The diffusion equations become
\begin{equation}
-\kappa^{2}C_{a}=\dfrac{\partial^{2}C_{a}}{\partial\xi^{2}},\:\:\:\mathrm{0\leq\xi\leq1/2},\label{eq:175-1}
\end{equation}
\begin{equation}
-\kappa^{2}C_{b}=\dfrac{\partial^{2}C_{b}}{\partial\xi^{2}},\:\:\:\mathrm{1/2\leq\xi\leq1,}\label{eq:176-1}
\end{equation}
with the boundary conditions
\[
\dfrac{dC_{a}}{d\xi}(0)=0,\:\:\:\dfrac{dC_{b}}{d\xi}(1)=0.
\]
The solutions are
\[
C_{a}(\xi)=F\cos\xi\kappa,\:\:\:C_{b}(\xi)=G\cos\kappa(\xi-1),
\]
where $F$ and $G$ are constants. Inserting them back in Eqs.(\ref{eq:177})-(\ref{eq:180}),
the boundary conditions at $\xi=1/2$ become
\[
2\kappa^{2}X_{GB}+V-c_{v}^{*}\kappa\sin(\kappa/2)(F-G)+\lambda_{t}\cos(\kappa/2)(F-G)=0,
\]
\[
V+c_{v}^{*}\kappa\sin(\kappa/2)(F+G)=0,
\]
\[
2V+\rho\cos(\kappa/2)(F+G)=0,
\]
\[
2\kappa^{2}X_{GB}+V-2\lambda\cos(\kappa/2)(F-G)=0.
\]
Treating $X_{GB}$, $V$, $(F+G)$ and $(F-G)$ as unknowns, we have
a homogeneous linear system of equations
\begin{equation}
\begin{bmatrix}2\kappa^{2} &  & 1 &  & 0 &  &  & \lambda_{t}\cos(\kappa/2)-c_{v}^{*}\kappa\sin(\kappa/2)\\
0 &  & 1 &  & c_{v}^{*}\kappa\sin(\kappa/2) &  &  & 0\\
0 &  & 2 &  & \rho\cos(\kappa/2) &  &  & 0\\
2\kappa^{2} &  & 1 &  & 0 &  &  & -2\lambda\cos(\kappa/2)
\end{bmatrix}\begin{bmatrix}X_{GB}\\
V\\
F+G\\
F-G
\end{bmatrix}=\begin{bmatrix}0\\
0\\
0\\
0
\end{bmatrix}.\label{eq:Lin_Syst_1}
\end{equation}
This system of\lyxadded{Yuri Mishin}{Sat Nov  9 02:14:44 2019}{ equations}
has nontrivial solutions if its determinant is zero. This condition
gives the dispersion relation
\begin{equation}
2\kappa^{2}\left(\rho\cos(\kappa/2)-2c_{v}^{*}\kappa\sin(\kappa/2)\right)\left((2\lambda+\lambda_{t})\cos(\kappa/2)-c_{v}^{*}\kappa\sin(\kappa/2)\right)=0,\label{eq:Determinant}
\end{equation}
which factors into three branches that will be referred to as the
$\kappa=0$ branch, the $\kappa_{\rho}$ branch, and the $\kappa_{\lambda}$
branch.

On the $\kappa=0$ branch, the linear system (\ref{eq:Lin_Syst_1})
becomes
\begin{equation}
\begin{bmatrix}0 & 1 & 0 & \lambda_{t}\\
0 & 1 & 0 & 0\\
0 & 2 & \rho & 0\\
0 & 1 & 0 & -2\lambda
\end{bmatrix}\begin{bmatrix}X_{GB}\\
V\\
F+G\\
F-G
\end{bmatrix}=\begin{bmatrix}0\\
0\\
0\\
0
\end{bmatrix}.\label{eq:Lin_Syst_1-1}
\end{equation}
and has the trivial solution $F=G=V=0$ with arbitrary $X_{GB}$.
This branch reflects the fact that the equilibrium can be reached
for any GB position $X_{GB}$.

On the $\kappa_{\rho}$ branch, the root $\kappa_{\rho}$ of Eq.(\ref{eq:Determinant})
satisfies the condition
\[
\dfrac{\rho}{2c_{v}^{*}\kappa_{\rho}}=\tan\dfrac{\kappa_{\rho}}{2},
\]
which is a transcendental equation of the form
\begin{equation}
\dfrac{Z}{x}=\tan x\label{eq:transcendent-equation}
\end{equation}
with a constant $Z$. It has a well-known graphical interpretation
with the root given by the intersection of the left-hand side $y=Z/x$
and the right hand side $y=\tan x$. For positive $Z$, the smallest
root lies in the range $0<x<\pi/2$. On this branch, the parameter
$Z$ is
\begin{equation}
Z_{\rho}=\dfrac{\rho}{4c_{v}^{*}}.\label{eq:Z_rho}
\end{equation}
The linear system (\ref{eq:Lin_Syst_1}) takes the form
\begin{equation}
\begin{bmatrix}2\kappa_{\rho}^{2} &  & 1 &  & 0 &  & (\lambda_{t}-\rho/2)\cos(\kappa_{\rho}/2)\\
0 &  & 2 &  & \rho\cos(\kappa_{\rho}/2) &  & 0\\
0 &  & 2 &  & \rho\cos(\kappa_{\rho}/2) &  & 0\\
2\kappa_{\rho}^{2} &  & 1 &  & 0 &  & -2\lambda\cos(\kappa_{\rho}/2)
\end{bmatrix}\begin{bmatrix}X_{GB}\\
V\\
F+G\\
F-G
\end{bmatrix}=\begin{bmatrix}0\\
0\\
0\\
0
\end{bmatrix}\label{eq:Lin_Syst_1-3}
\end{equation}
and the corresponding normal mode is proportional to
\begin{equation}
\begin{bmatrix}X_{GB}\\
V\\
F+G\\
F-G
\end{bmatrix}=\begin{bmatrix}\rho\cos(\kappa_{\rho}/2)\\
-2\rho\kappa_{\rho}^{2}\cos(\kappa_{\rho}/2)\\
4\kappa_{\rho}^{2}\\
0
\end{bmatrix}.\label{eq:Lin_Syst_1-2}
\end{equation}
Since $F=G$, the vacancy concentration is continuous across the GB.
Note that in this mode, the GB migration velocity $d\xi_{GB}/d\tau-\eta/2=0$.
The boundary does not move relative to the mean velocity of the grains.
The only GB process occurring is the vacancy generation and absorption.

On the $\kappa_{\lambda}$ branch, the root $\kappa_{\lambda}$ of
Eq.(\ref{eq:Determinant}) satisfies the condition
\[
\dfrac{2\lambda+\lambda_{t}}{c_{v}^{*}\kappa_{\lambda}}=\tan\dfrac{\kappa_{\lambda}}{2},
\]
which is the same transcendent equation (\ref{eq:transcendent-equation})
with the constant
\begin{equation}
Z_{\lambda}=\dfrac{2\lambda+\lambda_{t}}{2c_{v}^{*}}.\label{eq:Z_lambda}
\end{equation}
The linear system becomes
\begin{equation}
\begin{bmatrix}2\kappa_{\lambda}^{2} &  & 1 &  & 0 &  &  & -2\lambda\cos(\kappa_{\lambda}/2)\\
0 &  & 1 &  & (2\lambda+\lambda_{t})\cos(\kappa_{\lambda}/2) &  &  & 0\\
0 &  & 2 &  & \rho\cos(\kappa_{\lambda}/2) &  &  & 0\\
2\kappa_{\lambda}^{2} &  & 1 &  & 0 &  &  & -2\lambda\cos(\kappa_{\lambda}/2)
\end{bmatrix}\begin{bmatrix}X_{GB}\\
V\\
F+G\\
F-G
\end{bmatrix}=\begin{bmatrix}0\\
0\\
0\\
0
\end{bmatrix}\label{eq:Lin_Syst_1-4}
\end{equation}
and the normal mode is proportional to
\begin{equation}
\begin{bmatrix}X_{GB}\\
V\\
F+G\\
F-G
\end{bmatrix}=\begin{bmatrix}\lambda\cos(\kappa_{\lambda}/2)\\
0\\
0\\
\kappa_{\lambda}^{2}
\end{bmatrix}.\label{eq:Lin_Syst_1-2-1}
\end{equation}
In this mode, there is no driving force for the vacancy creation or
absorption at the GB and thus no lattice flow ($V=0$). The boundary
can migrate and the vacancy concentration profile is discontinuous
at the boundary, while its spatial derivative remains continuous ($F=-G$).

\newpage\clearpage{}

\renewcommand{\thetable}{\arabic{table}}{}%
\begin{table}
\noindent \begin{centering}
{}\caption{Summary of model
parameters for the numerical study of mixed-mode vacancy relaxation.\label{tab:Parameters}}
\textbf{\vfill{}
}\textbf{}%
\begin{tabular}[t]{ccllllllllll>{\raggedright}m{0.4\textwidth}}
\hline 
Case &  & $\lambda_{t}$ &  & $\lambda$ &  & $\rho$ &  & $Z_{\rho}$ &  & $Z_{\lambda}$ &  & GB properties\tabularnewline
\hline 
1 &  & $300$ &  & $4\times10^{4}$ &  & $10^{10}$ &  & $5.0\times10^{13}$ &  & $8.03\times10^{8}$ &  & Mobile, efficient
vacancy sink/source, weak diffusion barrier.\tabularnewline
2 &  & $300$ &  & $10^{-6}$ &  & $10^{10}$ &  & $5.0\times10^{13}$ &  & $3.0\times10^{6}$ &  & Sluggish, efficient
vacancy sink/source, weak diffusion barrier.\tabularnewline
3 &  & $300$ &  & $4\times10^{4}$ &  & $10^{-6}$ &  & $5.0\times10^{-3}$ &  & $8.03\times10^{8}$ &  & Mobile, poor vacancy
sink/source, weak diffusion barrier.\tabularnewline
4 &  & $300$ &  & $10^{-6}$ &  & $10^{-6}$ &  & $5.0\times10^{-3}$ &  & $3.0\times10^{6}$ &  & Sluggish, poor vacancy
sink/source, weak diffusion barrier.\tabularnewline
5 &  & $10^{-3}$ &  & $4\times10^{4}$ &  & $10^{10}$ &  & $5.0\times10^{13}$ &  & $8.03\times10^{8}$ &  & Mobile, efficient
vacancy sink/source, strong diffusion barrier.\tabularnewline
6 &  & $10^{-3}$ &  & $10^{-6}$ &  & $10^{10}$ &  & $5.0\times10^{13}$ &  & $10.02$ &  & Sluggish, efficient
vacancy sink/source, strong diffusion barrier.\tabularnewline
7 &  & $10^{-3}$ &  & $4\times10^{4}$ &  & $10^{-6}$ &  & $5.0\times10^{-3}$ &  & $8.0\times10^{8}$ &  & Mobile, poor vacancy
sink/source, strong diffusion barrier.\tabularnewline
8 &  & $10^{-3}$ &  & $10^{-6}$ &  & $10^{-6}$ &  & $5.0\times10^{-3}$ &  & $10.02$ &  & Sluggish, poor vacancy
sink/source, strong diffusion barrier.\tabularnewline
\hline 
\end{tabular}
\par\end{centering}

\end{table}

\newpage\clearpage{}

{}

\end{document}